\documentclass[twoside,12pt]{article}
\usepackage{epsfig}

\newcommand{\be}{\begin{equation}}
\newcommand{\ee}{\end{equation}}
\newcommand{\bea}{\begin{eqnarray}}
\newcommand{\eea}{\end{eqnarray}}

\begin{document}

\title{Equation of state for dense supernova matter}

\author{Ch.C. Moustakidis\\
$^{}$ Department of Theoretical Physics, Aristotle University of
Thessaloniki, \\ 54124 Thessaloniki, Greece }

\maketitle

\begin{abstract}
We provide an equation of state for high density supernova matter
by applying a momentum-dependent effective interaction. We focus
on the study of the equation of state of high-density and
high-temperature nuclear matter containing leptons (electrons and
neutrinos) under the chemical equilibrium condition. The
conditions of charge neutrality and equilibrium under
$\beta$-decay process lead first to the evaluation of the lepton
fractions and afterwards the evaluation of internal energy,
pressure, entropy  and in total to the equation of state of hot
nuclear matter for various isothermal cases. Thermal effects on
the properties and equation of state of nuclear matter are
evaluated and analyzed in the framework of the proposed effective
interaction model. Since supernova matter is characterized by a
constant entropy we also present  the thermodynamic properties for
isentropic case. Special attention is dedicated to the study of
the contribution of the components of $\beta$-stable nuclear
matter to the entropy per particle, a quantity of great interest
for the study of structure and collapse of supernova.
\vspace{0.3cm}

PACS number(s): 21.65.+f, 21.30.Fe, 24.10.Pa, 26.60.+c, 26.50.+x,
26.60.Kp \\

Keywards: Hot Nuclear Matter, Effective Interaction, Equation of
State, Nuclear Symmetry Energy, Proton Fraction, Supernova,
Neutron Star.
\end{abstract}

\section{Introduction}

Knowledge of the properties of the equation of state (EOS) of hot
asymmetric nuclear matter is of fundamental importance to
understand the physical mechanism of the iron core collapse of a
massive star which produces a type-II supernova, and the rapid
cooling of a new born hot neutron star. Additionally, the EOS
defines the chemical composition, both qualitative and
quantitative, of the hot nuclear matter.
\cite{Bethe-90,Prakash-97,Shapiro-04,Bombaci-94}. Supernova
explosions and neutron stars provide a unique laboratory where the
EOS of nuclear matter can be investigated. A great opportunity to
explore the EOS and properties of dense neutron-rich matter is
available at the accelerator facility at GSI, and in future, will
be also available through the high-energy radioactive beams at the
planned Facility for Rare Isotope Accelerator (FRIA)
\cite{Li-01,Danielewicz-02}.

There is a wealth of existing literature regarding the EOS of
supernova matter
\cite{Bethe-90,Bombaci-94,Eid-77,Lattimer-78,Bethe-79,Antia-80,Eid-80,
Barranco-81,Brown-82,Bowers-82,Cooperstein-84,Baron-85-1,Baron-85-2,Cooperstein-85,Lattimer-85,Pi-86,Brown-88,Muller-90,
Lattimer-91,Das-92,Cooperstein-93,Mayle-93,Takatsuka-94,Sumiyoshi-94,Onsi-94,Onsi-97,Das-07}.
Supernova matter which exists in a collapsing supernova core and
eventually forms a hot neutron star at birth is another form of
nuclear matter distinguished in the participation of degenerate
neutrinos and electrons \cite{Takatsuka-94}. It is characterized
by almost constant entropy per baryon $S=1-2$ (in units of the
Boltzmann constant $k_B$) throughout the density $n$ and also by a
high and almost constant lepton fraction $Y_l=0.3-0.4$ in contrast
with ordinary neutron star matter where $S=0$ and $Y_l \leq 0.05$.
These characteristics are caused by the effects of
neutrino-trapping which occurs in the dense supernova core where a
neutron star is formed.

This paper is a continuation of our previous work concerning the
EOS of hot $\beta$-stable nuclear matter in cases where neutrinos
have left the system \cite{Moustakidis-08-2}. More specifically,
in order to study the properties and the EOS of hot nuclear
matter,  a momentum-dependent effective interaction model (MDIM)
has been applied, one which is able to reproduce the results of
more microscopic calculations of dense matter at zero temperature
and which can be extended to finite temperature
\cite{Prakash-97,Moustakidis-08-2,Moustakidis-07,Moustakidis-08-1}.
 The main incentive for the present study is the fact that only
few calculations of the equation of state of the supernova matter
at high densities are available, although  at lower densities
($n<n_0$) (where $n_0=0.16$ fm$^{-3}$ is the saturation density)
reliable results are already available. For our purposes, we have
applied a model which, in comparison to previous models, has the
specific property that the temperature affects not only the
kinetic, but  also the interaction part of the energy density . In
this way, we are able to simultaneously study thermal effects on
the kinetic part of the symmetry energy and symmetry free energy,
in addition to the interaction part of the above quantities
\cite{Moustakidis-07}. This is significant in the sense that the
density dependent behavior of the symmetry energy and symmetry
free energy strongly influence  the values of the proton fraction
and as a consequence the composition of  hot $\beta$-stable
nuclear matter.

Our focus of interest is on the study of  dense supernova matter.
It has been  speculated that matter at densities up to about
$n=4n_0$ may be present in the core collapse of type-II supernova
\cite{Bethe-90}. The present work can also be applied to the study
of a neutron star at its birth, which is of particular interest as
such a star creates a new form of matter under extreme conditions.
In particular, proto-neutron stars are identified as a final stage
of a supernova collapse. At this stage, a proto-neutron star is
hot and composed of the so-called supernova matter.

In addition, we examine the two findings of the previous work of
Takatsuka et al. \cite{Takatsuka-94,Takatsuka-96} concerning
supernova matter. The first one is concerned with the finding that
the population of the components is remarkably constant both in
baryon density $n$ and temperature $T$ and
 the proton fraction $Y_p$ is very large (e.g.,
$Y_p\simeq 0.3$ for $Y_l=0.4$) in contrast with that of neutron
star matter. The second one concerns the finding that the EOS of
dense supernova matter is by far stiffer than that of neutron star
matter and correspondingly, hot neutron stars at birth are not
only "fat" but hot as well compared to usual cold neutron stars.
We broaden our study further by examining the influences of the
temperature on the stiffness of EOS compared to the cold case.

The article is organized as follows. In section II the model and
relative formulas are discussed and analyzed. Results are reported
and discussed in section III, whereas the summary of the work is
given in section IV.

\section{The model}

The model we use here, which has already been presented and
analyzed in our previous papers
\cite{Moustakidis-07-a,Moustakidis-08-2,Moustakidis-07,Moustakidis-08-1},
is designed to reproduce the results of the microscopic
calculations of both nuclear and neutron-rich matter at zero
temperature and can be extended to finite temperature
\cite{Prakash-97}. We provide the main characteristics of the
model as follows:

The energy density of the asymmetric nuclear matter (ANM) is given
by the relation
\begin{equation}
\epsilon(n_n,n_p,T)=\epsilon_{kin}^{n}(n_n,T)+\epsilon_{kin}^{p}(n_p,T)+
V_{int}(n_n,n_p,T), \label{E-D-1}
\end{equation}
where $n_n$ ($n_p$) is the neutron (proton) density and the total
baryon density is $n=n_n+n_p$. The contribution of the kinetic
parts are
\begin{equation}
\epsilon_{kin}^n(n_n,T)+\epsilon_{kin}^p(n_p,T)=2 \int \frac{d^3
k}{(2 \pi)^3}\frac{\hbar^2 k^2}{2m} \left(f_n(n_n,k,T)+
f_p(n_p,k,T) \right), \label{E-K-D-1}
\end{equation}
where $f_{\tau}$, (for $\tau=n,p$) is the Fermi-Dirac distribution
function.

Including the effect of finite-range forces between nucleons, the
potential contribution is parameterized as follows
\cite{Prakash-97}
\begin{eqnarray}
V_{int}(n_n,n_p,T)&=&\frac{1}{3}An_0\left[\frac{3}{2}-(\frac{1}{2}+x_0)I^2\right]u^2
+\frac{\frac{2}{3}Bn_0\left[\frac{3}{2}-(\frac{1}{2}+x_3)I^2\right]u^{\sigma+1}}
{1+\frac{2}{3}B'\left[\frac{3}{2}-(\frac{1}{2}+x_3)I^2\right]u^{\sigma-1}}
\nonumber \\ &+& u \sum_{i=1,2}\left[C_i \left({\cal J}_n^i+{\cal
J}_p^i\right)  + \frac{(C_i-8Z_i)}{5}I\left({\cal J}^i_n-{\cal
J}_p^i\right)\right], \label{V-all}
\end{eqnarray}
where
\begin{equation}
{\cal J}_{\tau}^i= \ 2 \int \frac{d^3k}{(2\pi)^3}
g(k,\Lambda_i)f_{\tau}(n_{\tau},k,T). \label{J-tau}
\end{equation}

In Eq.~(\ref{V-all}), $I$ is the asymmetry parameter
($I=(n_n-n_p)/n$) and $u=n/n_0$, with $n_0$ denoting the
equilibrium symmetric nuclear matter density, $n_0=0.16$
fm$^{-3}$. The asymmetry parameter $I$ is related to the proton
fraction $Y_p$ by the equation $I=(1-2Y_p)$.  The parameters $A$,
$B$, $\sigma$, $C_1$, $C_2$ and $B'$ which appear in the
description of symmetric nuclear matter are determined in order
that $E(n=n_0)-mc^2=-16$ {\rm MeV}, $n_0=0.16$ fm$^{-3}$, and the
incompressibility are $K=240$ {\rm MeV}.  The additional
parameters $x_0$, $x_3$, $Z_1$, and $Z_2$ used to determine the
properties of asymmetric nuclear matter are treated as parameters
constrained by empirical knowledge \cite{Prakash-97}.

The first two terms of the right-hand side of Eq.~(\ref{V-all})
arise from local contact nuclear interaction which lead to power
density contributions as in the standard Skyrme equation of state.
The first one corresponds to attractive interaction and the second
one to repulsive. They are assumed to be temperature independent.
The third term describes the effects of finite range interactions
according to the choice of the function $g(k,\Lambda_i)$, and is
the temperature dependent part of the interaction \cite{Li-01}.
This interaction is attractive and important at low momentum, but
it weakens and disappears at very high momentum.  The function,
$g(k,\Lambda_i)$, suitably chosen to simulate finite range
effects, has the following form
\begin{equation}
 g(k,\Lambda_i)=\left[1+\left(\frac{k}{\Lambda_{i}}\right)^2
\right]^{-1}, \label{g-1} \end{equation}
where the finite range parameters are $\Lambda_1=1.5 k_F^{0}$ and
$\Lambda_2=3 k_F^{0}$ and $k_F^0$ is the Fermi momentum at the
saturation point $n_0$.

The energy density of asymmetric nuclear matter at density $n$ and
temperature $T$, in a good approximation, is expressed as
\begin{equation}
\epsilon(n,T,I)=\epsilon(n,T,I=0)+\epsilon_{sym}(n,T,I),
\label{e-asm-1}
\end{equation}
where
\begin{equation}
\epsilon_{sym}(n,T,I)=nI^2 E_{sym}^{tot}(n,T) =n I^2
\left(E_{sym}^{kin}(n,T)+E_{sym}^{int}(n,T)\right).
\label{e-sym-1}
\end{equation}
In Eq.~(\ref{e-sym-1}) the nuclear symmetry energy
$E_{sym}^{tot}(n,T)$ is separated into two parts corresponding to
the kinetic contribution $E_{sym}^{kin}(n,T)$ and the interaction
contribution $E_{sym}^{int}(n,T)$.

From Eqs.~(\ref{e-asm-1}) and (\ref{e-sym-1}) and setting $I=1$,
we find that the nuclear symmetry energy $E_{sym}^{tot}(n,T)$ is
given by
\begin{equation}
E_{sym}^{tot}(n,T)=\frac{1}{n}\left(\epsilon(n,T,I=1)-\epsilon(n,T,I=0)
\right). \label{Esym-d-1}
\end{equation}
Thus, from Eq.~(\ref{Esym-d-1}) and by a suitable choice of the
parameters $x_0$, $x_3$, $Z_1$ and $Z_2$, we can obtain different
forms for the density dependence of the symmetry energy
$E_{sym}^{tot}(n,T)$.

It is well known that the need to explore different forms for
$E_{sym}^{tot}(n,T)$ stems from the uncertain behavior at high
density \cite{Prakash-97}. The high-density behavior of symmetry
energy is the worst known property of dense matter
\cite{Kutschera-94,Li-02,Fuchs-06}, with different nuclear models
giving contradictory predictions. Recently, the density dependence
of the symmetry energy in the equation of state of isospin
asymmetric nuclear matter has been studied using isoscaling of the
fragment yields and the antisymmetrized molecular dynamic
calculation \cite{Shetty-07}. It was observed that the
experimental data at low densities are  consistent with the form
of symmetry energy, $E_{sym}(u)\approx 31.6u^{0.69}$, in close
agreement with those predicted by the results of variational
many-body calculations. In Ref.~\cite{Shetty-07} it was also
suggested that the heavy ion studies favor a dependence of the
form $E_{sym}(u)\approx 31.6u^{\gamma}$, where $\gamma=0.6-1.05$.
This constrains the form of the density dependence of the symmetry
energy at higher densities, ruling out an extremely "stiff" and
"soft" dependence \cite{Shetty-07}.

In a previous study conducted with isospin dependent
Boltzmann-Uehling-Uhlenbeck transport calculations, Chen et
al.~\cite{Chen-05} have shown that a stiff density dependence of
the symmetry energy parameterized as $E_{sym}(u)\approx
31.6u^{1.05}$ clearly explains the isospin diffusion data
\cite{Tsang-04} from NSCL-MSU (National Superconducting Cyclotron
Laboratory at Michigan State University).

In the present work, since we are primarily interested  in the
study of thermal effects on the nuclear symmetry energy and free
energy, we choose a specific form for this, enabling us to
accurately reproduce
  the results of many other theoretical studies
\cite{Lee-98,Sammarruca-08}. In Ref.~\cite{Lee-98} the authors
carried out a systematic analysis of the nuclear symmetry energy
in the formalism of the relativistic Dirac-Brueckner-Hartree-Fock
approach using the Bonn one-boson-exchange potential. In a very
recent work, \cite{Sammarruca-08} the authors applied a similar
method as in Ref.~\cite{Lee-98} for the microscopic predictions of
the equation of state in asymmetric nuclear matter. In this case
$E_{sym}(u)$ is obtained with the simple parametrization
$E_{sym}(u)=C u^{\gamma}$ with $\gamma=0.7-1.0$ and $C\approx 32$
MeV.  The authors concluded that a value of $\gamma$ close to
$0.8$ gives a reasonable description of their predictions although
the use of different functions in different density regions may be
best for an optimal fit \cite{Sammarruca-08}. The results of
Refs.~\cite{Lee-98,Sammarruca-08}  are well reproduced by
parameterizing the nuclear symmetry energy according to the
following formula
\begin{equation}
E_{sym}^{tot}(n,T=0)= 13 u^{2/3}+17 F(u),\label{Esym-3}
\end{equation}
where the first term of the right part of Eq.~(\ref{Esym-3})
corresponds to the contribution of the kinetic energy and the
second term to the contribution of the interaction energy.

For the function $F(u)$, which parameterizes the interaction part
of the symmetry energy, we apply the following form
\begin{equation}
F(u)=u. \label{Fu-form}
\end{equation}
The parameters $x_0$, $x_3$, $Z_1$ and $Z_2$ are chosen so that
Eq.~(\ref{Esym-d-1}), for $T=0$ reproduces the results of
Eq.~(\ref{Esym-3}) for  the function $F(u)=u$.

In general, in order to obtain different forms for the density
dependence of $E_{sym}(n)$, the function $F(u)$  can be
parameterized as follows \cite{Prakash-97}

\begin{equation}
F(u)=\sqrt{u},\qquad  F(u)=u, \qquad F(u)=2u^2/(1+u). \label{fu-1}
\end{equation}

It is worthwhile to point out that the above parametrization of
the interacting part of the nuclear symmetry energy is extensively
used for the study of neutron star properties
\cite{Prakash-97,Prakash-94} as well as  the study of the
collisions of neutron-rich heavy ions at intermediate energies
\cite{Li-97,Baran-05}. For a very recent review of the
applications of the proposed momentum dependent effective
interaction model and the specific parametrization of it see
Ref.~\cite{Li-08} (and references therein).

\subsection{Thermodynamic description of hot nuclear matter}
In order to study the properties of nuclear matter at finite
temperature, we need to introduce the Helmholtz free energy $F$
which is written as \cite{Goodstein-85,Fetter-03}

\begin{equation}
F(n,T,I)=E(n,T,I)-TS(n,T,I). \label{Free-1}
\end{equation}
In Eq.~(\ref{Free-1}),  $E$ is the internal energy per particle,
$E=\epsilon/n$, and $S$ is the entropy per particle, $S=s/n$. From
Eq.~(\ref{Free-1}) it is also concluded that for $T=0$, the free
energy $F$ and the internal energy $E$ coincide.

The entropy density $s$ has the same functional form as that of a
non-interacting gas system, given by the equation
\begin{equation}
s_{\tau}(n,T,I)=-2\int \frac{d^3k}{(2\pi)^3}\left[f_{\tau} \ln
f_{\tau}+(1-f_{\tau}) \ln(1-f_{\tau})\right],  \label{s-den-1}
\end{equation}
while the pressure and the chemical potentials defined as follows
\cite{Goodstein-85,Fetter-03}
\begin{equation}
P=n^2\left(\frac{\partial \epsilon/n}{\partial n}\right)_{S,N_i},
\qquad \qquad \qquad \mu_{i}=\left(\frac{\partial
\epsilon}{\partial n_i}\right)_{S,V,n_{j\neq i}}. \label{P-m-E}
\end{equation}

At this point we shall examine the properties and the EOS of
nuclear matter by considering an isothermal process. In this case,
the pressure and the chemical potentials are related to  the
derivative of the total free energy density $f=F/n$. More
specifically, they are defined as follows
\begin{equation}
P=n^2\left(\frac{\partial f/n}{\partial n}\right)_{T,N_i}, \qquad
\qquad \mu_{i}=\left(\frac{\partial f}{\partial
n_i}\right)_{T,V,n_{j\neq i}}. \qquad \qquad \label{P-m-F}
\end{equation}

The pressure $P$ can also be  calculated  from the equation
\cite{Goodstein-85,Fetter-03}
\begin{equation}
P=Ts-\epsilon+\sum_{i}\mu_in_i. \label{P-1}
\end{equation}

It is also possible to calculate the entropy per particle $S(n,T)$
by differentiating the free energy density $f$ with respect to the
temperature, that is
\begin{equation}
S(n,T)=- \left(\frac{\partial f/n}{\partial T} \right)_{V,N_i}.
\label{S-dif-f}
\end{equation}

The comparison of the two entropies, that is from
Eqs.~(\ref{s-den-1}) and (\ref{S-dif-f}), provides a test of the
approximation used in the present work (see
Ref.~\cite{Moustakidis-08-2}).
It is easy to demonstrate by applying Eq.~(\ref{P-m-F}) that (see
for a proof \cite{Prakash-94} as well as \cite{Burgio-07})
\begin{eqnarray}
\mu_n&=&F+u\left(\frac{\partial F}{\partial
u}\right)_{Y_p,T}-Y_p\left(\frac{\partial F}{\partial
Y_p}\right)_{n,T}, \nonumber
\\ \mu_p&=&\mu_n+\left(\frac{\partial F}{\partial
Y_p}\right)_{n,T}, \nonumber \\
\hat{\mu}&=&\mu_n-\mu_p=-\left(\frac{\partial F}{\partial
Y_p}\right)_{n,T}.
 \label{mu-p-n}
\end{eqnarray}

We can define the symmetry free energy per particle $F_{sym}(n,T)$
by the following parabolic approximation (see also
\cite{Burgio-07,Xu-07-3})
\begin{equation}
F(n,T,I)=F(n,T,I=0)+I^2F_{sym}(n,T)=
F(n,T,I=0)+(1-2Y_p)^2F_{sym}(n,T),\label{Free-Parabolic}
\end{equation}
where
\begin{equation} F_{sym}(n,T)= F(n,T,I=1)-F(n,T,I=0).
\label{Free-asym}
\end{equation}
It is worth noting that the above approximation is not valid from
the beginning, but one needs to check the validity of the
parabolic law in the present model before using it. In
Ref.~\cite{Moustakidis-08-2} we have proved the validity of the
approximation (\ref{Free-Parabolic}).

Now, by applying Eq.~(\ref{Free-Parabolic}) in Eq.~(\ref{mu-p-n}),
we obtain the  key relation
\begin{equation}
\hat{\mu}=\mu_n-\mu_p=4(1-2Y_p)F_{sym}(n,T). \label{mhat-2}
\end{equation}
The above equation is similar to that obtained for cold nuclear
matter by replacing $E_{sym}(n)$ with $F_{sym}(n,T)$.  The finding
that, for both quantities ($E(n,T,I)$ and $F(n,T,I)$), the
dependence of the asymmetry parameter $I$ can be approximated very
well by a quadratic dependence leads to the conclusion that the
entropy $S(n,T,I)$ must also exhibit quadratic dependence of $I$
that is
\begin{equation}
S(n,T,I)=S(n,T,I=0)+I^2S_{sym}(n,T) \label{S-sym-1}
\end{equation}
where
\begin{equation} S_{sym}(n,T)= S(n,T,I=1)-S(n,T,I=0)=\frac{1}{T}(E_{sym}(n,T)-F_{sym}(n,T)).
\label{S-sym-1}
\end{equation}
In order to check the validity of the parabolic approximation
(\ref{S-sym-1}), we plot in Fig.~1 the difference
$S(n,T,I=1)-S(n,T,I=0)$ as a function of $I^2$ at temperature
$T=10$ and $T=30$ MeV for two baryon densities, i.e., $n=0.2$
fm$^{-3}$ and $n=0.4$ fm$^{-3}$. It is thus evident that in a good
approximation, an almost linear relation holds between
$S(n,T,I=1)-S(n,T,I=0)$ and $I^2$, especially for low values of
$I^2$.

Also noteworthy in the present model is that due to temperature
dependence of the interaction part of the energy density, the
temperature affects both the kinetic part contribution on the
entropy $S$ and the interaction part. Hence, the total entropy per
baryon must be written as follow $S_{tot}=S_{kin}+S_{int}$.
\subsection{$\beta$-equilibrium in hot proto-neutron star and supernova}

Stable high density nuclear matter must be in chemical equilibrium
with all type of reactions, including the weak interactions in
which $\beta$ decay and electron capture take place simultaneously
\begin{equation}
n \longrightarrow p+e^{-}+\bar{\nu}_e, \qquad \qquad p +e^{-}
\longrightarrow n+ \nu_e. \label{b-reaction}
\end{equation}
Both types of reactions change the electron per nucleon fraction,
$Y_e$ and thus affect the equation of state. In a previous study,
we assumed that neutrinos generated in these reactions left the
system \cite{Moustakidis-08-2}. The absence of neutrino-trapping
has a dramatic effect on the equation of state and is the main
cause of  a significant reduction in the values of the proton
fraction $Y_p$ \cite{Takatsuka-94,Takatsuka-96}.

The chemical equilibrium of reactions (\ref{b-reaction}) can be
expressed in terms of the chemical potentials for the four species
\begin{equation}
\mu_n+\mu_{\nu_{e}}=\mu_p+\mu_e. \label{chem-pot-1}
\end{equation}

The charge neutrality condition provides us with the equation
\begin{equation}
Y_p=Y_e, \label{neut-con}
\end{equation}
while the total fraction of leptons is given by
\begin{equation}
Y_{l}=Y_e+Y_{\nu_e}. \label{lepton-nu-1}
\end{equation}

Now, according to Eqs.~(\ref{mhat-2}) and (\ref{chem-pot-1}) we
have

\begin{equation}
\mu_{e}-\mu_{\nu_e}=\mu_n-\mu_p=4\left(1-2Y_p\right)F_{sym}(n,T).
\label{chem-lep-2}
\end{equation}
Moreover, the leptons (electrons, muons and neutrinos) density is
given by the expression
\begin{equation}
n_l=\frac{g}{(2\pi)^3}\int \frac{ {\rm d}{\bf
k}}{1+\exp\left[\frac{\sqrt{\hbar^2k^2c^2+m_l^2c^4}-\mu_l}{T}
\right]}, \label{n-lepton-1}
\end{equation}
\\
where $g$ stands for  the spin degeneracy ($g=2$ for electrons and
muons and $g=1$ for neutrinos). One can  self-consistently solve
Eqs.~(\ref{neut-con}), (\ref{lepton-nu-1}), (\ref{chem-lep-2}) and
(\ref{n-lepton-1}) in order to calculate the proton fraction
$Y_p$($=Y_e$), the neutrino fractions $Y_{\nu_e}$, as well as the
electron chemical potential $\mu_e$ as a function of the baryon
density $n$ for various values of temperature $T$.

Depending on the form of the symmetry energy, muons can appear at
high density. Prakash \cite{Prakash-94} has shown that the more
rapidly $F(u)$ increases with density, the lower the density at
which muons appear. For example, with $F(u)=u$, muons appear at
$u\sim 4$, while with the choice $F(u)=\sqrt{u}$, muons cannot
appear till $u\sim 8$. However, the presence of muons has very
little effect on the electron lepton fractions (compared to the
case without the inclusion of muons) since $Y_{\mu}$ remains
extremely small ($\sim 10^{-4}$) over a wide range of densities
\cite{Prakash-94}. Thus, we do not include the contribution of
muons in our study.

The next step is to calculate the energy density and pressure of
leptons given by the following formulae
\begin{equation}
\epsilon_{l}(n_l,T)=\frac{g}{(2\pi)^3}\int \frac{\sqrt{\hbar^2 k^2
c^2+m_l^2c^4} \ {\rm d}{\bf
k}}{1+\exp\left[\frac{\sqrt{\hbar^2k^2c^2+m_l^2c^4}-\mu_l}{T}
\right]}, \label{e-lepton-1}
\end{equation}
\begin{equation}
P_l(n_l,T)=\frac{1}{3}\frac{g(\hbar c)^2}{(2\pi)^3}\int
\frac{1}{\sqrt{\hbar^2 k^2 c^2+m_l^2c^4}}\frac{\ k^2 \ {\rm d}{\bf
k}}{1+\exp\left[\frac{\sqrt{\hbar^2k^2c^2+m_l^2c^4}-\mu_l}{T}
\right]}. \label{P-lepton-1}
\end{equation}

The entropy density $s$ has the same functional form as that of a
non-interacting gas system, given by the equation
\begin{equation}
s_{l}(n,T,I)=-g\int \frac{d^3k}{(2\pi)^3}\left[f_{l} \ln
f_{l}+(1-f_{l}) \ln(1-f_{l})\right].  \label{s-den-lept}
\end{equation}

The equation of state of hot nuclear matter in $\beta$-equilibrium
(considering that it consists of neutrons, protons, electrons and
neutrinos) can be obtained by calculating the total energy density
$\epsilon_{tot}$ as well as the total pressure $P_{tot}$. The
total energy density is given by
\begin{equation}
\epsilon_{tot}(n,T,I)=\epsilon_b(n,T,I)+\sum_{l=e,\nu_e}\epsilon_l(n,T,I),
\label{e-de-1}
\end{equation}
where $\epsilon_b(n,T,I)$ and $\epsilon_l(n,T,I)$ are the
contributions of baryons and leptons respectively. The total
pressure is
\begin{equation}
P_{tot}(n,T,I)=P_b(n,T,I)+\sum_{l=e,\nu_e}P_l(n,T,I), \label{Pr-1}
\end{equation}
where $P_b(n,T,I)$ is the contribution of the baryons (see
Eq.~(\ref{P-1})) i.e.
\begin{equation}
P_b(n,T,I)=T\sum_{\tau=p,n}s_{\tau}(n,T,I)+\sum_{\tau=n,p}n_{\tau}\mu_{\tau}(n,T,I)-\epsilon_b(n,T,I),
\label{Pr-2}
\end{equation}
while $P_l(n,T,I)$ is the contribution of the leptons (see
Eq.~(\ref{P-lepton-1})). From Eqs.~(\ref{e-de-1}) and (\ref{Pr-1})
we can construct the isothermal curves for energy and pressure and
finally derive the isothermal behavior of the equation of state of
hot nuclear matter under $\beta$-equilibrium.

\section{Results and Discussions}
We calculate the equation of state of hot asymmetric nuclear
matter by applying a momentum dependent effective interaction
model describing the baryons interaction. We consider that nuclear
matter contains neutrons, protons, electrons and neutrinos under
$\beta$-equilibrium and charge neutrality. The key quantities in
our calculations are the proton fraction $Y_p$ and also the
asymmetry free energy defined in Eq.~(\ref{Free-asym}). It is
worth pointing out that since the supernova explosion itself is a
dynamic phenomenon, the chemical composition of matter changes
according to the evolution of the star all the time
\cite{Sumiyoshi-94}. During supernova explosion, the chemical
composition of matter reaches equilibrium not in the whole star
 but locally. In our present work we assume matter in the
chemical equilibrium for simplicity in order to analyze the
properties of hot neutron star and supernova matter.

The validity of the parabolic approximation (\ref{Free-Parabolic})
has already been tested and presented in our previous work
\cite{Moustakidis-08-2}. $F_{sym}(u,T)$, for various values of the
temperature $T$, was derived with a least-squares fit to the
numerical values according to Eq.~(\ref{Free-asym}) and has the
form \cite{Moustakidis-08-2}
\begin{eqnarray}
F_{sym}(u;T=0)&=& 13 u^{2/3}+17u \nonumber\\
F_{sym}(u;T=5)&=&3.653+28.018u-1.5126u^2+0.185u^3-0.010u^4,
\nonumber \\
F_{sym}(u;T=10)&=&5.995+26.157u-0.827u^2+0.068u^3-0.002u^4, \nonumber \\
F_{sym}(u;T=20)&=&13.200+21.267u+0.800u^2-0.193u^3+0.014u^4,
\nonumber \\
F_{sym}(u;T=30)&=&21.087+17.626u+1.645u^2-0.289u^3+0.018u^4.
 \label{Fsym-T-fit}
\end{eqnarray}
where the case with $T=0$, is included as well. In that case
$F_{sym}$ coincides with $E_{sym}$.

Firstly, in order to clarify the contribution of the three terms
of the potential energy density, we plot the terms as a function
of the baryon density, in Fig.~2(a). In that figure we have that
\begin{eqnarray}
V^A&=&\frac{1}{3}An_0\left[\frac{3}{2}-(\frac{1}{2}+x_0)I^2\right]u^2,                 \nonumber\\
V^B&=&\frac{\frac{2}{3}Bn_0\left[\frac{3}{2}-(\frac{1}{2}+x_3)I^2\right]u^{\sigma+1}}
{1+\frac{2}{3}B'\left[\frac{3}{2}-(\frac{1}{2}+x_3)I^2\right]u^{\sigma-1}},                      \\
V^C&=& u \sum_{i=1,2}\left[C_i \left({\cal J}_n^i+{\cal
J}_p^i\right)  + \frac{(C_i-8Z_i)}{5}I\left({\cal J}^i_n-{\cal
J}_p^i\right)\right].                  \nonumber \label{3-terms}
\end{eqnarray}
The first term $V^A$ corresponds to attractive interaction where
the second $V^B$ corresponds  to repulsive interaction and
dominates for high values of  $n$ ($n>0.6$ fm$^{-3}$). Both of
the above terms are temperature independent. The third term $V^C$
contains the momentum dependent part of the interaction,
corresponds to attractive interaction and its main contribution is
to compete the repulsive interaction of $V^B$ for high values of
$n$ and as a consequence  avoid acausal behavior of the EOS  at
high densities. The term $V^C$ consists of two finite range terms,
one corresponding to a long-range attraction and the other to a
short-range repulsion.

Thermal effects on the momentum dependent term $V^C$ are displayed
in Fig.~2(b). The contribution of $V^C$ is plotted for various
values of $T$. It is concluded that thermal effects are more
pronounced for high values of $T$ ($T>10$ MeV) leading to a less
attractive contribution. More precisely, we find that for small
values of $n$ (i.e. $n=0.15$ fm$^{-3}$ ) $V^C$ increases (compared
to the cold case $T=0$) $3\%-20\%$ for $T=10-30$. For higher
values of $n$ the increase is even less.

The outline of our calculations procedure approach is the
following: For a fixed value of baryon density density $n$, lepton
fraction $Y_l$, temperature $T$ and initial trial value of proton
fraction $Y_p$ ($=Y_e$), Eq.~(\ref{n-lepton-1}) is solved in order
to calculate the chemical potential $\mu_e$. The knowledge of
$\mu_e$ allows the calculation of $\mu_{\nu_e}$ from
Eq.~(\ref{chem-lep-2}) which may then be used to infer the
neutrino fraction $Y_{\nu_e}$ from Eq.~(\ref{n-lepton-1}).
Finally, a new value of proton fraction $Y_p$ ($=Y_e$) is taken
from equation $Y_e=Y_l-Y_{\nu_e}$ and the procedure is repeated
from the beginning until a convergence is achieved.

In Fig.~3 we plot the fraction of electrons $Y_e$ and neutrinos
$Y_{\nu_e}$ versus the baryon density  $n$ for lepton fraction
$Y_l=0.3$ and $Y_l=0.4$ and for various values of T. It is
concluded that thermal effects are important, both for electron
and neutrinos fractions for low values of the baryon density $n$
i.e. $n<0.4$ fm$^{-3}$. $Y_e$ is an increasing function of $T$ and
consequently $Y_{\nu_e}$ is a decreasing function of $T$. For
higher values of $n$, the thermal effects on lepton's fraction are
unimportant.

At this point, following the discussion of Takatsuka et al.
\cite{Takatsuka-94}, we attempt to extend the discussion
concerning the dependence of equilibrium  fraction $Y_e$($= Y_p$)
on the baryon density as well as on the nuclear symmetry energy.
We ignore the temperature effect to clarify the situation.
Actually, the situation does not  change by including finite
temperature effects. The energy per baryon of supernova matter
$E_{SM}$ and cold neutron star matter $E_{NS}$ are expressed as
function of $n$ and $Y_p$ (see also ref.~\cite{Takatsuka-94}) as
\begin{eqnarray}
E_{SM}(n,Y_p)&=&E_b(n,Y_p)+E_e(n,Y_p)+E_{\nu_e}(n,Y_p) \\
&=&E_b(n,Y_p=0.5)+E_{sym}(n)(1-2Y_p)^2+253.6
u^{1/3}Y_p^{4/3}+319.516u^{1/3}(Y_l-Y_p)^{4/3}\nonumber \\
E_{NS}(n,Y_p)&=&E_b(n,Y_p)+E_e(n,Y_p) \\
&=&E_b(n,Y_p=0.5)+E_{sym}(n)(1-2Y_p)^2+253.6
u^{1/3}Y_p^{4/3}\nonumber \label{E-NSM}
\end{eqnarray}
where the symmetry energy $E_{sym}(n)$ is parameterized according
to Eq.~(\ref{fu-1}). $E_{sym}(n)$ is plotted in Fig. 4(a) for the
three different parametrizations. In the same figure we have
included recent results provided in reference~\cite{Sammarruca-08}
achieved by performing microscopic calculations in asymmetric
nuclear matter. In this case $E_{sym}(n)$ is obtained with the
simple parametrization
\[E_{sym}(u)=C u^{\gamma}\]
with $\gamma=0.8$ and $C=32$ MeV. It is obvious that the results
of the above parametrization, correspond very well with the
parametrization $F(u)=u$ which is proposed here.

The equilibrium proton fraction $Y_p$ is calculated by solving the
equation $\partial E_{SM,NS}/\partial Y_p=0$ for various values of
the density $n$, $E_{sym}(n)$ and $Y_l=0.4$ for supernova matter.
The results are presented in Fig.~4(b). In the case of cold
neutron star matter, $Y_p$ depends strongly on both  the baryon
density and  the values of the $E_{sym}(n)$. This is not the case
for supernova matter where the effect of nuclear symmetry energy
in determining $Y_p$ is less important than in cold neutron star
matter. In addition, $Y_p$, for a fixed parametrization of $F(u)$
is almost constant with respect to $n$.

Fig.~5 displays thermal effects on the chemical potential of
leptons for $Y_l=0.3$ and $Y_l=0.4$. In all cases, $\mu_l$ is a
slightly decreasing  function of T. In fact, the important
quantity for our calculations is the difference
$\hat{\mu}=\mu_e-\mu_{\nu_e}$  which is strictly constrained from
Eq.~(\ref{chem-lep-2}). So, it is appropriate to check the
validity of Eq.~(\ref{Free-Parabolic}) at least for proton
fraction $Y_p\approx 0.3$ (or $I^2\approx 0.16$). We found in our
previous work \cite{Moustakidis-08-2} (Fig.~1), that
Eq.~(\ref{Free-Parabolic}) is accurate for the values of the
proton fraction which are under consideration in the present work.
The use of the formula (\ref{chem-lep-2}) is very useful since it
can greatly simplify the coupled equations used for the
construction of the EOS. We mention here that to our knowledge,
the above treatment has never been applied for the study of the
EOS of supernova matter and  has been applied for the first time
in the present work.

In Fig.~6 we plot the contribution of the baryons $S_b$, leptons
$S_l$ and the total $S_{tot}$ to the entropy per baryon. In all
cases, $S$ is a decreasing function of the baryon density $n$.
Temperature affects appreciably both baryon and lepton
contribution. It should be noted that the contribution of baryons
$S_b$ may be written as $S_b=S_{kin}+S_{int}$, where the term
$S_{kin}$  originates from  the temperature effect on the kinetic
part of the energy density and $S_{int}$ reflects thermal effects
on the potential energy density. More precisely, entropy density
$s$, according to equation (\ref{s-den-1}), is an increasing
function of the diffuseness of the Fermi-Dirac distribution
$f_{\tau}(n,k,T)$. As indicated in our previous work
\cite{Moustakidis-08-2} (Fig.~11), the effect of the diffuseness
of the distribution $f_{\tau}(n,k,T)$ is pronounced for low values
of the baryon density $n$ and for high values of $T$. But as we
have shown above, thermal effect on the momentum dependent term
$V^C$ is important for low values of $n$ and also high values of
$T$. Therefore, we conclude that the therm $V^C$ has a more
pronounced effect on the entropy density $s$ mainly for low values
of $n$. For higher values of $n$ the contribution of $V^C$ on $s$
is less important.

Furthermore, the lepton contribution $S_l$ is an increasing
function of the lepton fraction $Y_l$, while the baryon
contribution is almost independent by $Y_l$. For the electron and
neutrino entropy density, our present results can be reproduced
with good accuracy, at least for low values of $T$, by applying
the analytical formula used  by Onsi {\it et al.}
\cite{Onsi-94,Onsi-97}
\begin{equation}
s_e=\frac{1}{3}\frac{\mu_e^2}{(\hbar c)^3} T, \qquad \qquad
\mu_e=\hbar c(3\pi^2 Y_e n)^{1/3}. \label{Se-Onsi}
\end{equation}
\begin{equation}
s_{\nu_e}=\frac{1}{6}\frac{\mu_{\nu_e}^2}{(\hbar c)^3} T, \qquad
\qquad \mu_{\nu_e}=\hbar c(6\pi^2 Y_{\nu_e} n)^{1/3}.
\label{Sne-Onsi}
\end{equation}
According to the above formula, the specific contribution of the
leptons (electrons and neutrinos) to the entropy per baryon has
the form
\begin{equation}
S_{e,\nu_e}=s_{e,\nu_e}/n \sim
\left(\frac{Y_{e,\nu_e}^2}{n}\right)^{1/3}T. \label{Se-Onsi-2}
\end{equation}
Eq.~(\ref{Se-Onsi-2}) gives an nice explanation for the density
and temperature dependence of $S_l$ presented in Fig.~6(b).

In Fig.~7, we display the contribution to internal energy $E$ from
baryons $E_b$ and leptons $E_l$  for $Y_l=0.3$ and $Y_l=0.4$ and
for various values of $T$. The most striking aspect is that the
lepton energy, $E_l=E_e+E_{\nu_e}$, dominates in the internal
energy of the matter up to $n\sim 0.6$ fm$^{-3}$ (for $Y_l=0.3$)
and $n\sim 0.8$ fm$^{-3}$ (for $Y_l=0.4$). This is a
characteristic of the supernova matter and is in remarkable
contrast with the situation of cold neutron star matter
\cite{Takatsuka-94}. The contribution from baryon $E_b$ gets
larger with the increase of $n$ and is comparable with $E_l$ for
high values of $n$.

It is worth pointing out that the above situation depends on the
combination of the stiffness of nuclear equation of state (values
of incompressibility and density dependent behavior of the nuclear
symmetry energy ) and the lepton fraction. Nonetheless, the main
feature is unaltered, especially up to low values of $n$
\cite{Takatsuka-94}. Moreover, the contribution on the lepton
energy, as is presented in Fig.~8,  originates mainly from
electrons while neutrino contribution is smaller (but not
negligible).

The present results for the electron and neutrino energy per
baryon, can also be accurately reproduced, at least for low values
of $T$, by applying the analytical formula used by Onsi {\it et
al.} \cite{Onsi-94,Onsi-97} where the energy density $\epsilon_l$
and energy per baryon $E_l$ of the leptons are given by
\begin{equation}
\epsilon_e=\frac{1}{4\pi^2}\frac{\mu_{e}^4}{(\hbar
c)^3}\left(1+\frac{2}{3}\frac{\pi^2T^2}{\mu_{e}^2}\right), \qquad
E_e\sim (Y_e^4 n)^{1/3}\left(1+{\cal C} \frac{T^2}{(Y_e
n)^{2/3}}\right) \label{en-den-ele}
\end{equation}
\begin{equation}
\epsilon_{\nu_e}=\frac{1}{8\pi^2}\frac{\mu_{\nu_e}^4}{(\hbar
c)^3}\left(1+\frac{2}{3}\frac{\pi^2T^2}{\mu_{\nu_e}^2}\right) ,
\qquad E_{\nu_e}\sim (Y_{\nu_e}^4 n)^{1/3}\left(1+\tilde{{\cal C}}
\frac{T^2}{(Y_{\nu_e} n)^{2/3}}\right). \label{en-den-neut}
\end{equation}
In Eqs.~(\ref{en-den-ele}) and (\ref{en-den-neut}), ${\cal C}$ and
$\tilde{\cal C}$ are constants.

The contributions of baryon and leptons on the total pressure are
presented in Fig.~9. In contrast to the situation of the internal
energy, the nuclear part contribution plays a more important role
compared with the lepton part. The lepton pressure $P_l$ is
comparable to baryon pressure $P_b$ up to $n\sim 0.2$ fm$^{-3}$,
but for higher values of $n$ it is significantly small. What is
more, the main part of $P_l$ originates from electrons compared to
neutrinos as presented in Fig.~10.

As pointed out by Bethe et al. \cite{Bethe-79}, the crucial
feature in determining the evaluation of a collapsing
pre-supernova core is that the entropy per particle is very low,
of the order of unity (in units of the Boltzmann constant $k_B$),
and nearly constant during all the stages of the collapse up to
the shock wave formation. Therefore, the collapse is an adiabatic
process of a highly ordered system. So, since the supernova matter
is characterized by a constant entropy and constant lepton
fraction, we shall also discuss the properties under this
condition. This can be done by converting the results for
isothermal case ($T$=const) into those for adiabatic case
($S$=const) in terms of the $T-n$ relation constrained by a
constant entropy.

The $T=T(n)$ relation is constructed by $\{T,n\}$ values to
satisfy $S(n,T)$=const in an $S-n$ diagram. Fig.~11 shows the
results for $Y_l=0.3$ and $Y_l=0.4$ for $S=1$. Temperature is an
increasing function of $n$. Furthermore, for the same density, the
temperature is higher for lower values of $Y_l$. The values of $T$
for various values of $n$ are derived, for the two cases, with the
least-squares fit method and found to take the general form
\[T(n)=an^{b},\]
where $a=35.412$, $b=0.70481$ for $Y_l=0.3$ and $a=32.35706$,
$b=0.67694$ for $Y_l=0.4$. The results of this study are in
 very good agreement with those of Takatsuka et al.
\cite{Takatsuka-94}. The stars at lower density denote the
$\{T,n\}$ values for $S=1$ and $Y_l=0.4$ which are derived from
Lattimer et al. \cite{Lattimer-85}. It is concluded that the
temperature increases considerably when moving from the outer part
of the star to the center in order to maintain a constant value of
the entropy per baryon.

By applying the relation $T-n$ presented in Fig.~11, the fractions
$Y_i$ of isothermal case (see Fig.~3) is converted into the
isentropic one for $S=1$. The population of constituents is
plotted in Fig.~12 as a function of $n$ for $Y_l=0.3$ and
$Y_l=0.4$. The most striking feature of the results is the slight
dependence of the fraction $Y_i$ from the baryon density $n$ (the
same behavior and similar results have been found in
Ref.~\cite{Takatsuka-94}). To sum up, during the adiabatic
collapse of a supernova, the population of the constituents
(neutrons, protons, electrons and neutrinos) are almost the same
independent of the density $n$.

The entropy contributions (for $S=1$) from the constituents are
presented in Fig.~13. The contributions of the baryon are
increasing functions of $n$, the contribution of electrons is
increasing function of $n$, while $S_{\nu_e}$ is almost
independent of density. Roughly, neutrons, protons, electrons and
neutrinos contribute to $S$ ($=1$) by about $50, 30, 17, 3 \%$
(for $Y_l=0.3$) and $45, 30, 20, 5 \%$ (for $Y_l=0.4$)
respectively. Our results are quite consistent with those of
Ref.~\cite{Takatsuka-94}.

Fig.~14, displays the internal energies per baryon of respective
components $E_i$ versus $n$ for $Y_l=0.3$, $Y_l=0.4$ and $S=1$.
The main conclusions of the results are similar to those of the
isothermal case (see Fig.~7). The lepton energy (which mainly
originated from electrons contributions) dominates in the internal
energy of the supernova matter even for high values of the density
$n$. The nuclear contribution on the internal energy dominates
only in  high values of $n$ (depending on the lepton fraction
$Y_l$).

Finally, in Fig.~15 we compare the EOS's between supernova matter
and cold neutron star matter. The case for supernova matter
corresponds to $S=1$ and $Y_l=0.3$. It is thus clear that the
internal energy $E_{tot}$ of supernova matter (SM) is remarkably
larger than that of neutron star matter (NS). As far as the
nucleon part $E_b$ is concerned, the $E_b$ in SM is slightly lower
than that in NS due to the large energy gain in symmetry energy
(see also \cite{Takatsuka-94}). However, the lepton contribution
on the internal energy $E_l$ is remarkably larger in SN matter
compared to NS matter due to the effect of a large lepton
fraction, that is, a large kinetic energy of abundant leptons.
High temperature also contributes to the stiffening, but it is
less effective than the high lepton fractions (see also Fig.~7).
The present results also correspond well with those presented by
Takatsuka et al. \cite{Takatsuka-94,Takatsuka-96} a few years ago

\section{Summary}
The evaluation of the equation of state of hot nuclear matter is a
major challenge for nuclear physics and astrophysics. EOS is the
basic ingredient necessary for studying the supernova explosion as
well as for determining the properties of  hot neutron stars.  The
motive for the present work has been to apply a momentum-dependent
interaction model for the study of the hot nuclear matter EOS
under $\beta$-equilibrium. We have calculated the lepton fractions
by applying the constraints for chemical equilibrium and charge
neutrality. The internal energy and also the pressure have been
calculated as  functions of baryon density and for various values
of temperature. Special attention has been dedicated to the study
of the contribution of the components of $\beta$-stable nuclear
matter on the entropy per particle, a quantity of great interest
in the study of structure and collapse of supernova. We have
presented and analyzed the contribution of each component.
Finally, we have presented the EOS of $\beta$-stable hot nuclear
matter, by taking into account and analyzing  the contributions to
the total pressure of each component. The above EOS can be applied
to the evaluation of the gross properties of hot neutron stars
i.e. mass and radius.


\section*{Acknowledgments}
The author  would like to thank Professor Tatsuyauki Takatsuka for
valuable comments and correspondence.



\newpage
\begin{figure}
\centering
\includegraphics[height=8.0cm,width=8cm]{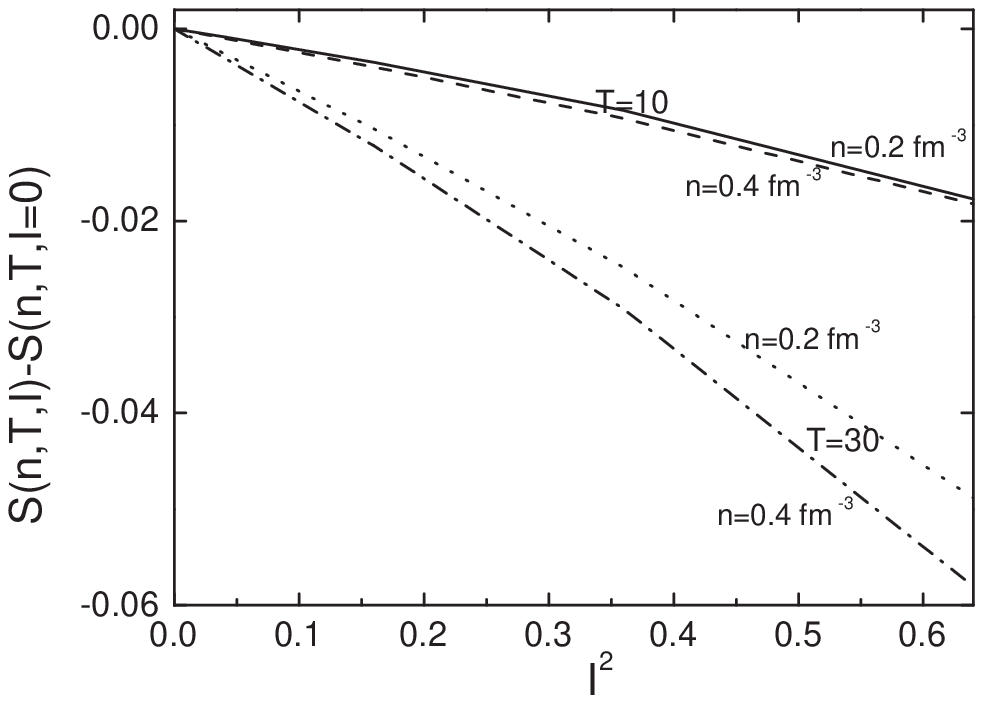}
\caption{The difference $S(n,T,I)-S(n,T,I=0)$ as a function of
$I^2$ at temperature $T=10$ MeV and $T=30$ MeV, for two baryon
densities.} \label{}
\end{figure}
\begin{figure}
\vspace{1cm} \centering
\includegraphics[height=8.0cm,width=8cm]{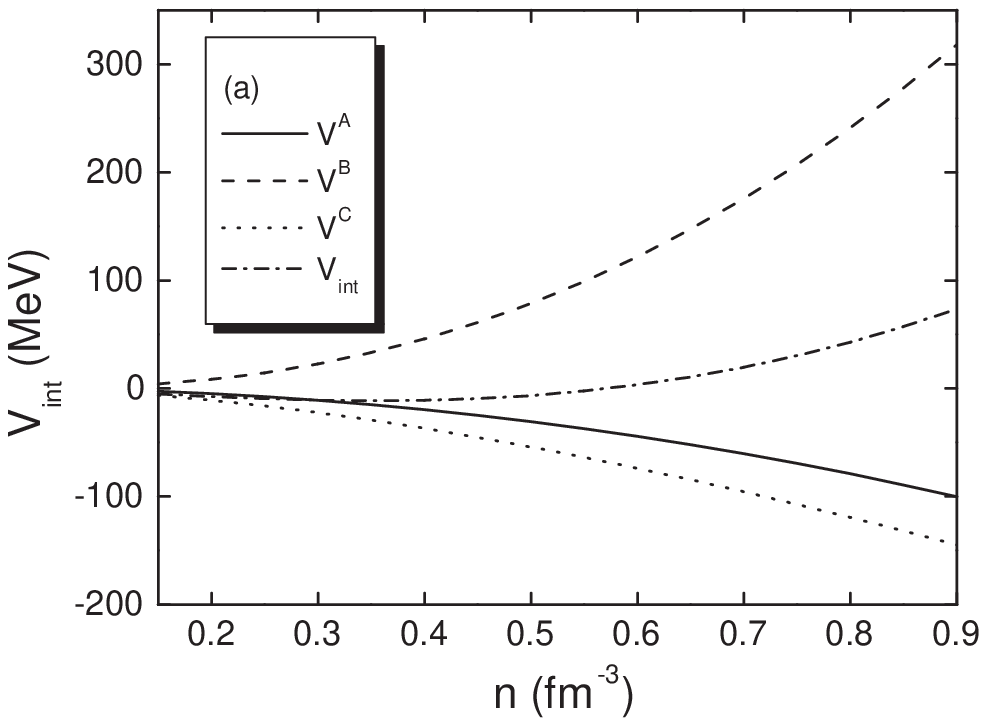}
\includegraphics[height=8.0cm,width=8cm]{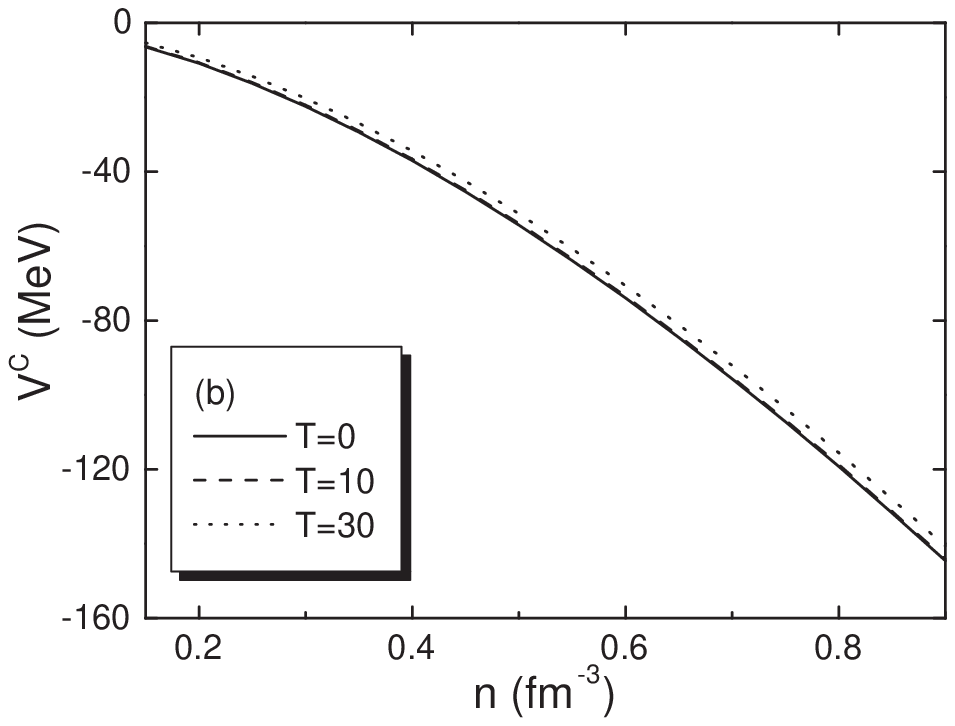}
\caption{a) The contribution of the various terms $V^A$, $V^B$,
$V^C$ and the total potential energy density $V_{int}$ as a
function of the baryon density  b) The momentum dependent term
$V^C$ as a function of the baryon density at temperature $T=0, 10,
30$ MeV. } \label{}
\end{figure}
\begin{figure}
\vspace{2cm} \centering
\includegraphics[height=8.0cm,width=8cm]{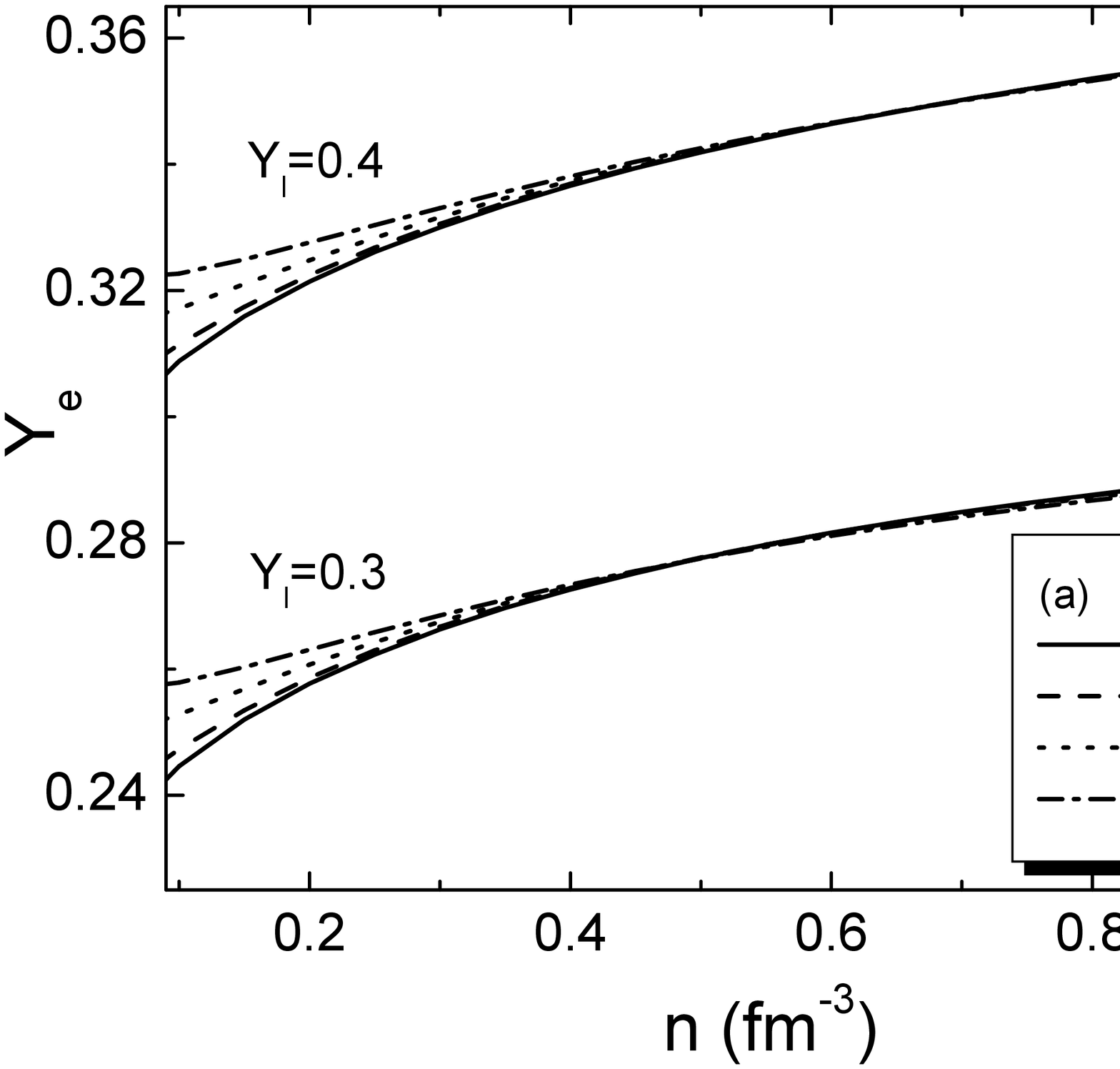}
\includegraphics[height=8.0cm,width=8cm]{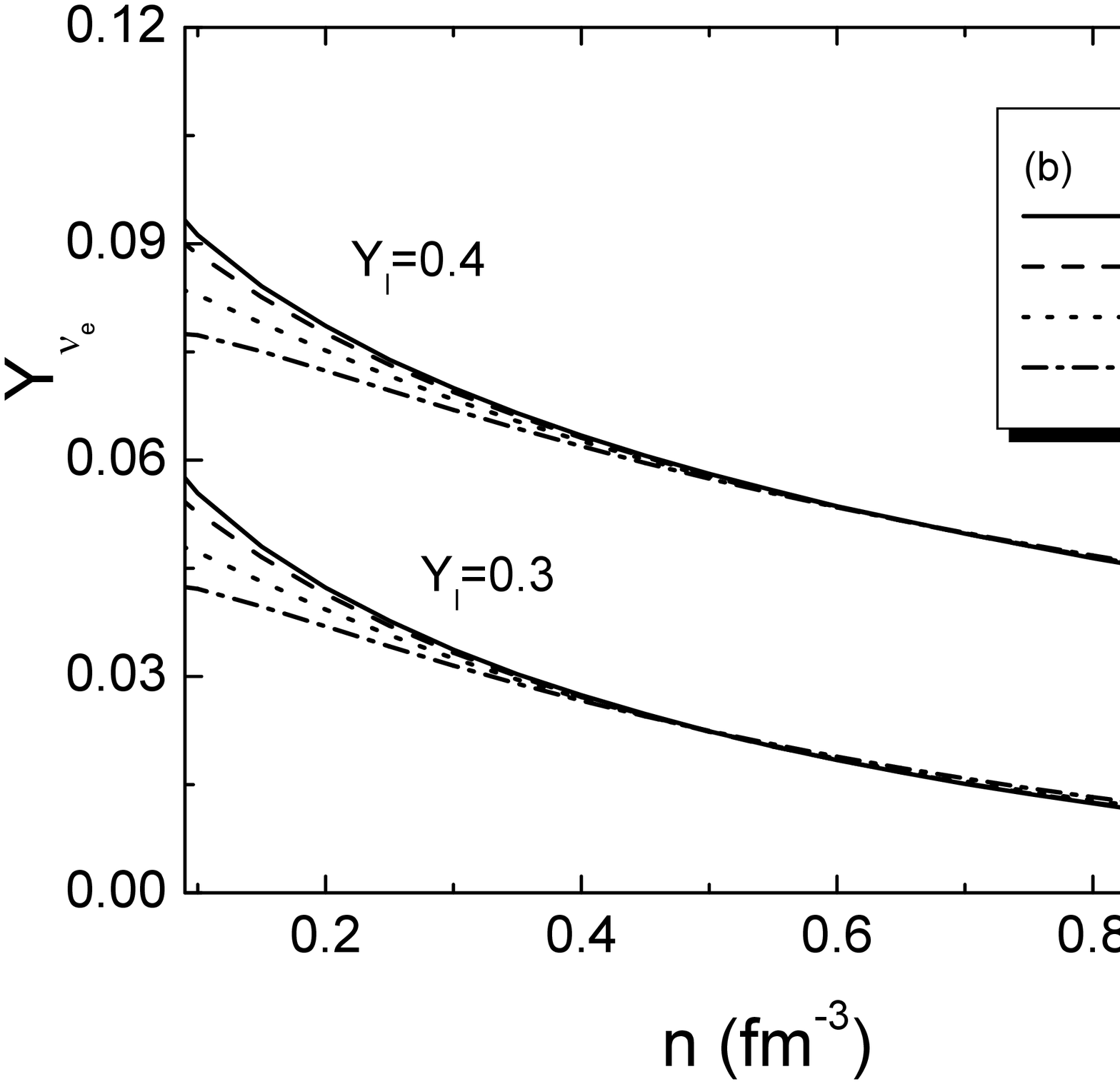}
\vspace{-3cm} \caption{Electron $Y_e$ and neutrino $Y_{\nu_e}$
fraction versus baryon density for various values of T for total
lepton fraction a) $Y_l=0.3$ and b) $Y_l=0.4$. }  \label{}
\end{figure}
\begin{figure}
\centering
\includegraphics[height=7.8cm,width=7.8cm]{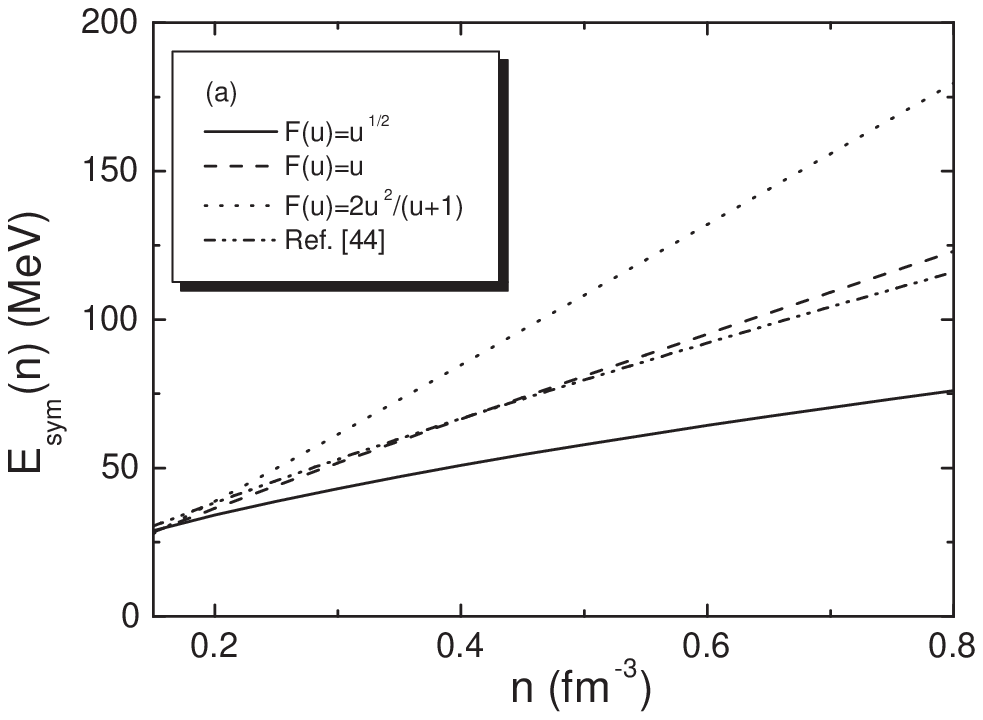}
\includegraphics[height=8.0cm,width=8cm]{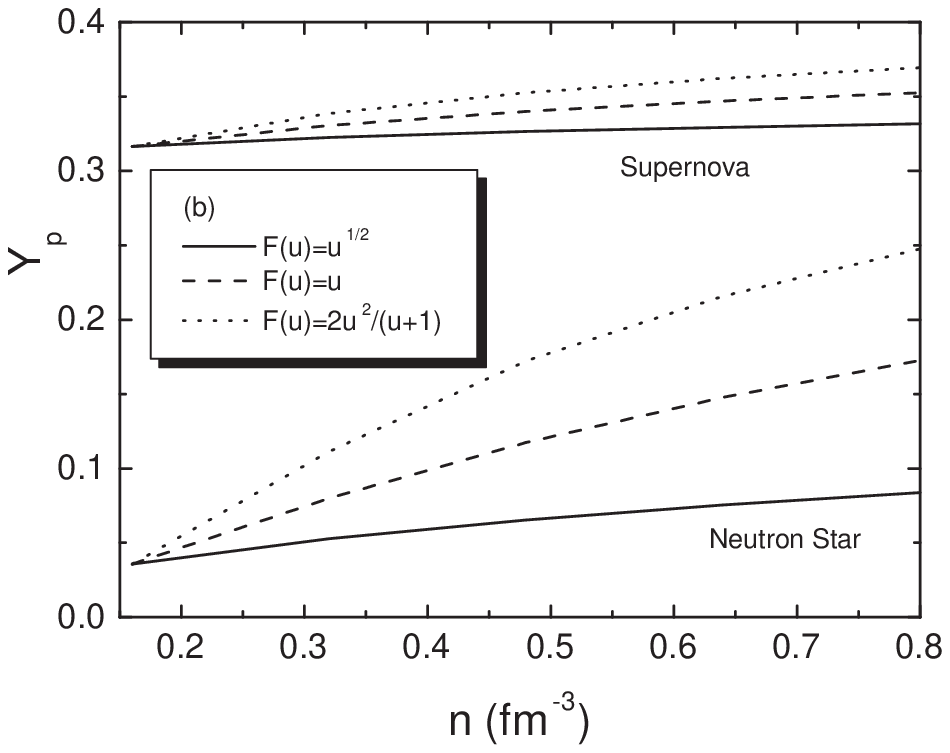}
\caption{a) The nuclear symmetry energy for three different
parametrization (equation~(\ref{fu-1})) of the interaction part
with the results of reference~\cite{Sammarruca-08} (see text for
more details) and b) the proton fraction $Y_p$ versus baryon
density for cold neutron star matter (down curves) and supernova
matter (up curves) for the three different parametrization of the
nuclear symmetry energy. } \label{}
\end{figure}
\begin{figure}
\vspace{1cm} \centering
\includegraphics[height=8.0cm,width=8cm]{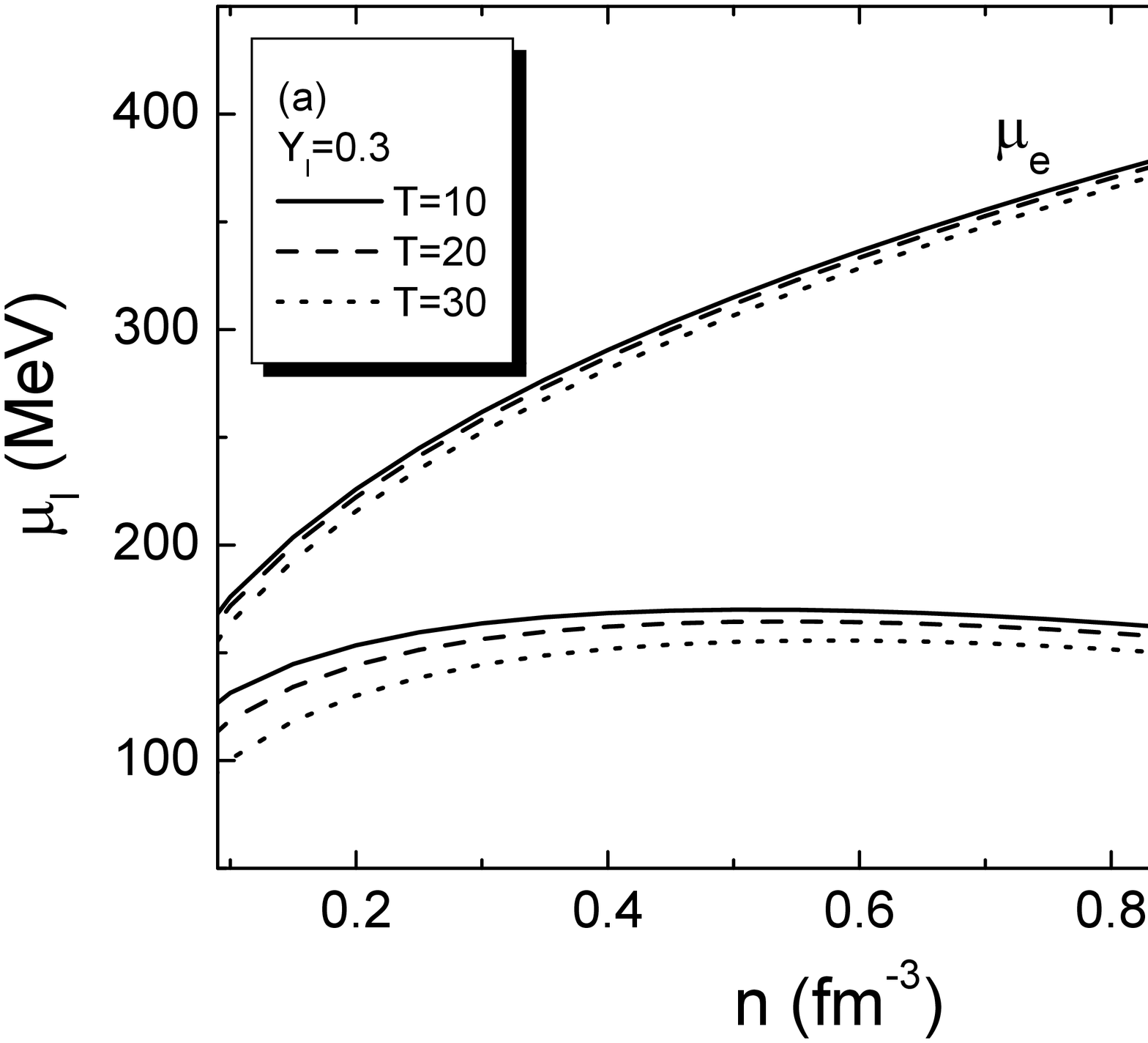}
\includegraphics[height=8.0cm,width=8cm]{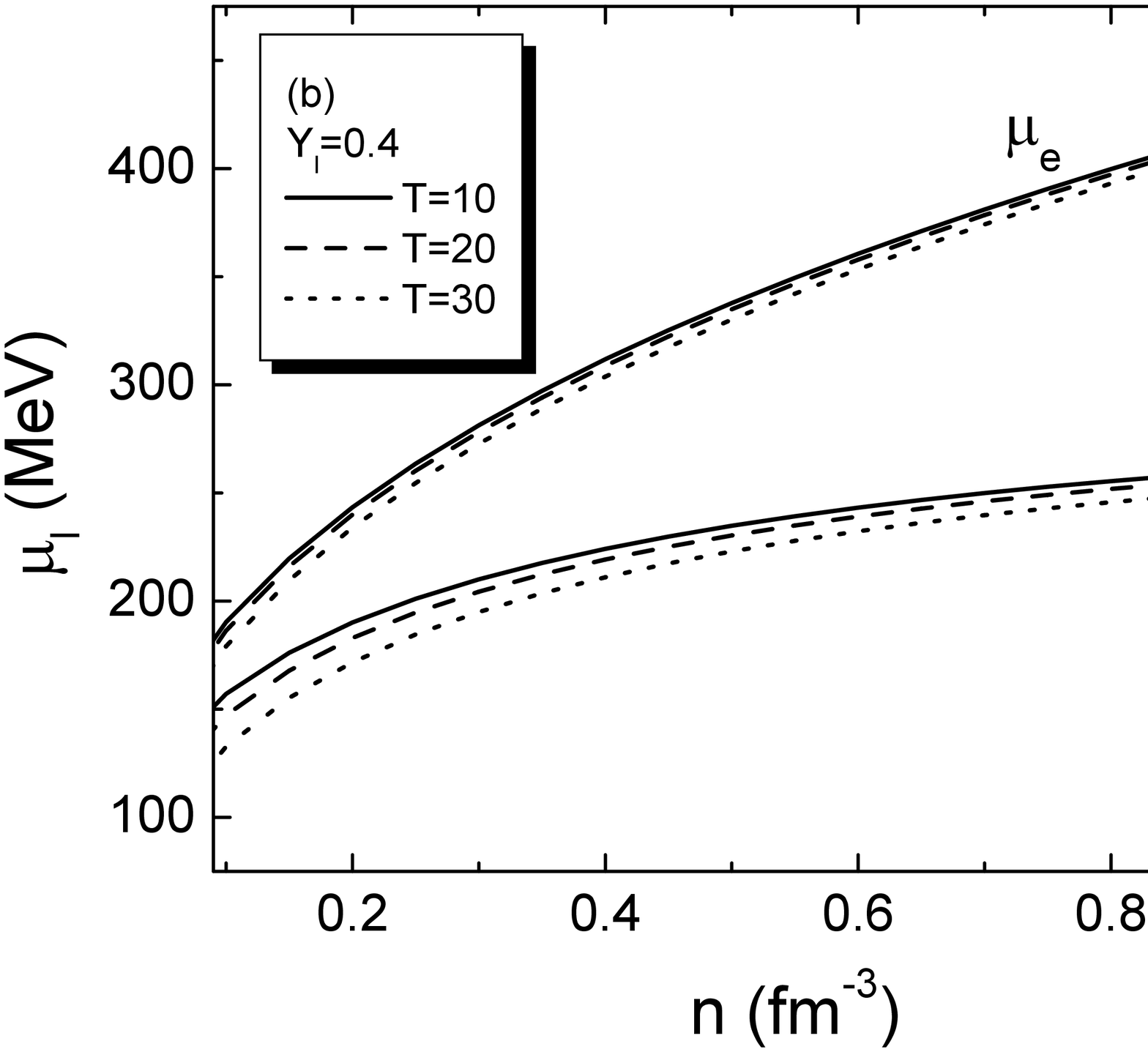}
\vspace{-3cm} \caption{Electron $\mu_e$ and neutrino $\mu_{\nu_e}$
chemical potentials  versus baryon density for various values of T
for total lepton fraction a) $Y_l=0.3$ and b) $Y_l=0.4$. }
\label{}
\end{figure}
\begin{figure}
\centering
\includegraphics[height=8.0cm,width=8cm]{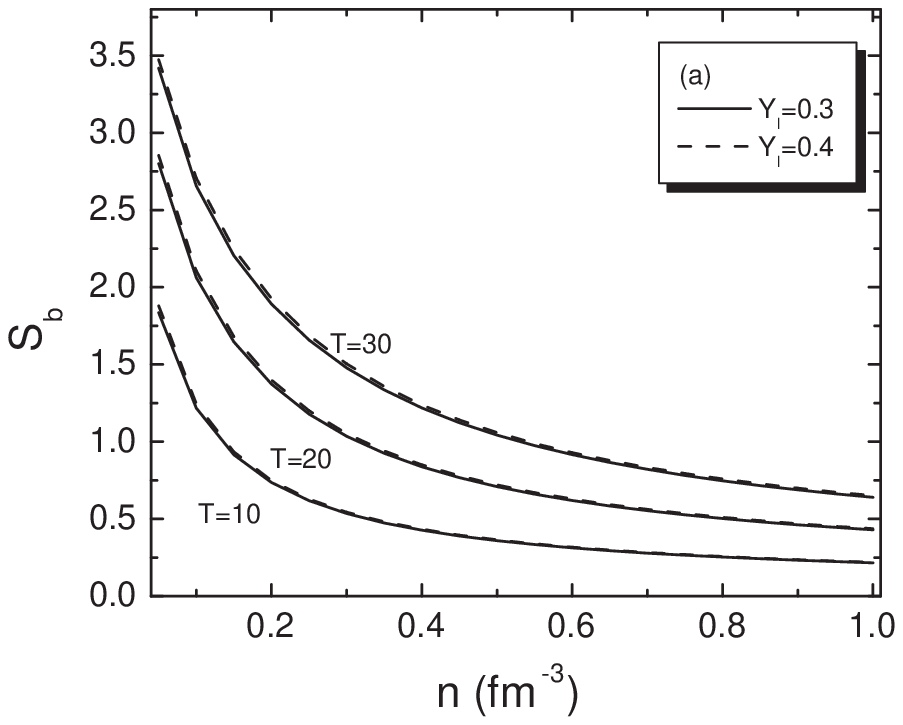}
\includegraphics[height=8.0cm,width=8cm]{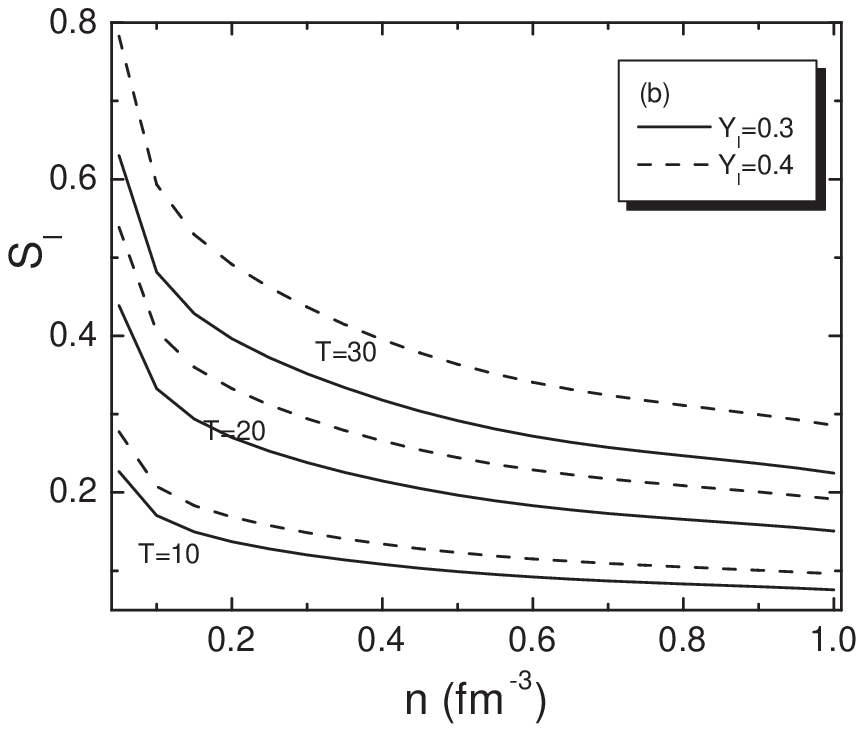}
\includegraphics[height=8.0cm,width=8cm]{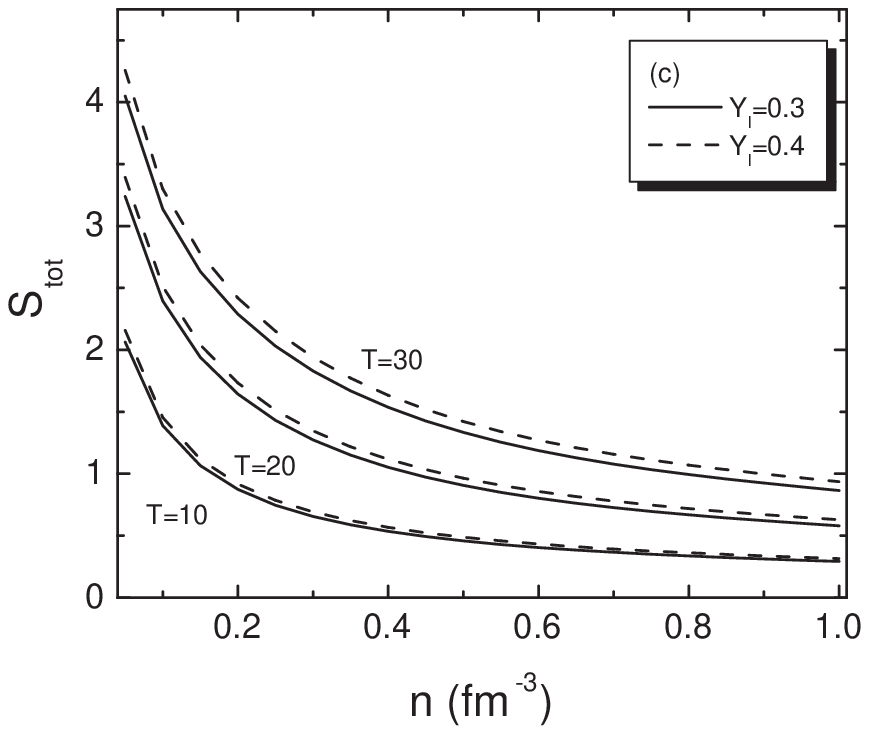}
\caption{Contribution to the total entropy per particle of a)
baryons ($S_b$), b) leptons ($S_l$) and c) the total entropy
($S_{tot}$) versus the baryon density for various values of T for
total lepton fraction $Y_l=0.3$ and $Y_l=0.4$. } \label{}
\end{figure}
\begin{figure}
\vspace{2cm} \centering
\includegraphics[height=8.3cm,width=8.3cm]{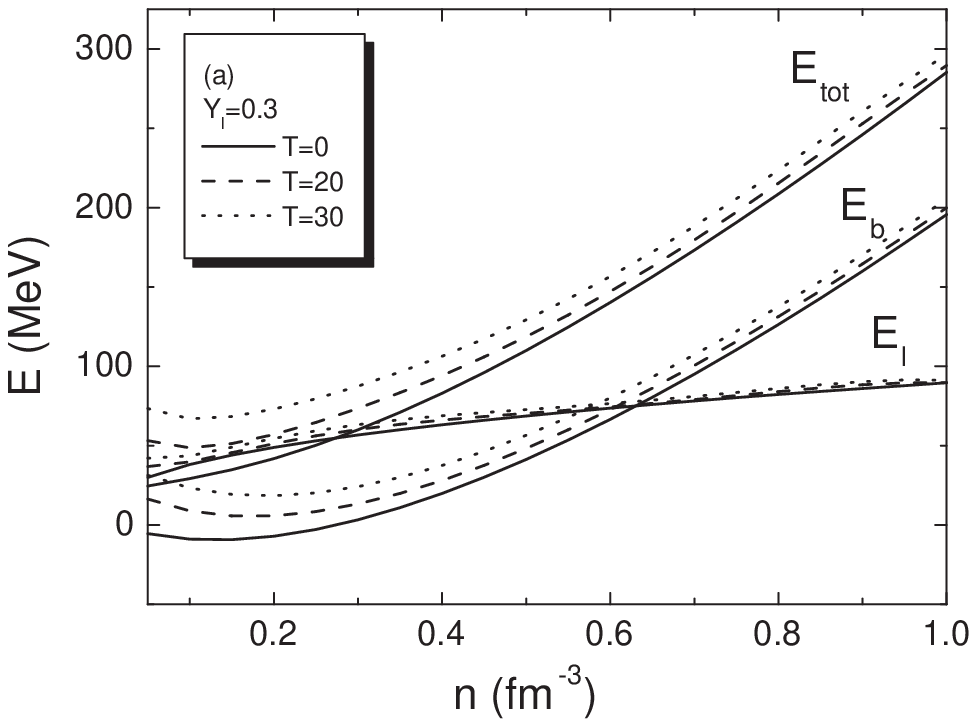}
\includegraphics[height=8.0cm,width=8.0cm]{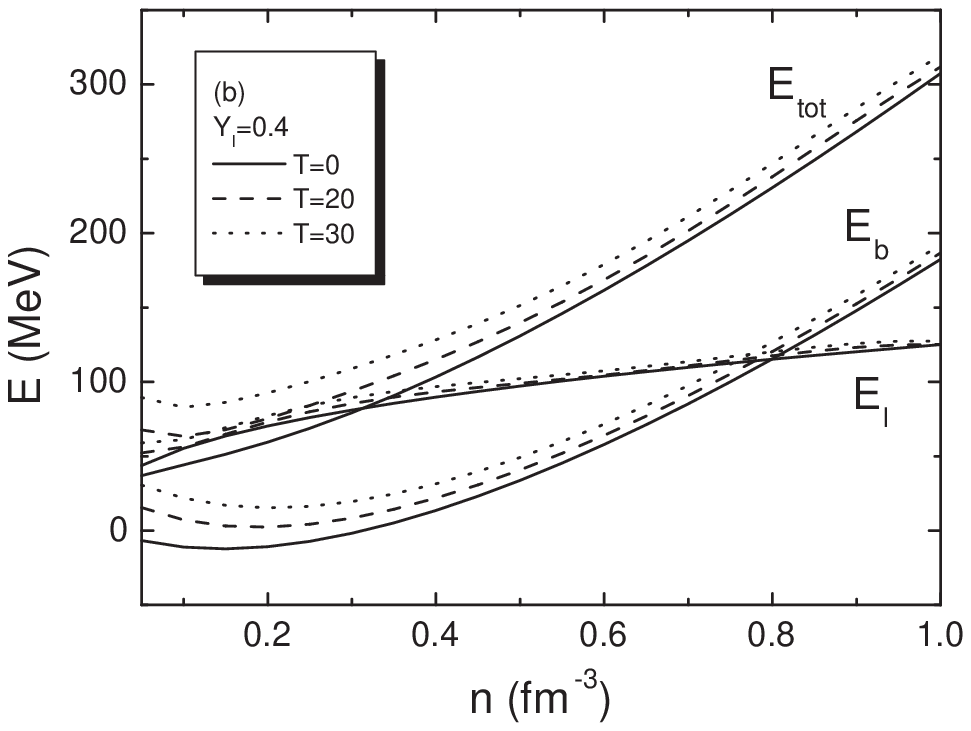}
\vspace{0.3cm} \caption{Contribution to the total energy per
particle of baryons ($E_b$), leptons ($E_l$) and  the total energy
($E_{tot}$) versus the baryon density for various values of T for
total lepton fraction  a) $Y_l=0.3$ and b) $Y_l=0.4$.} \label{}
\end{figure}
\begin{figure}
\centering
\includegraphics[height=8.0cm,width=8cm]{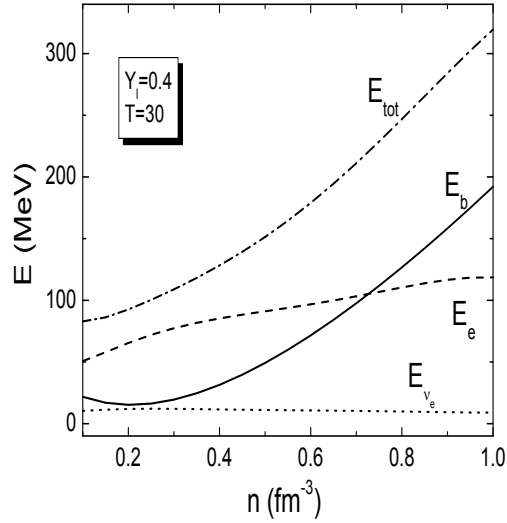}
\vspace{-3cm}
\caption{Energies per particle of respective components $E_i$
versus the baryon density $n$ for the specific case with $T=30$
MeV and $Y_l=0.4$. }  \label{}
\end{figure}
\vspace{5cm}
\begin{figure}
\vspace{2cm} \centering
\includegraphics[height=8.0cm,width=8cm]{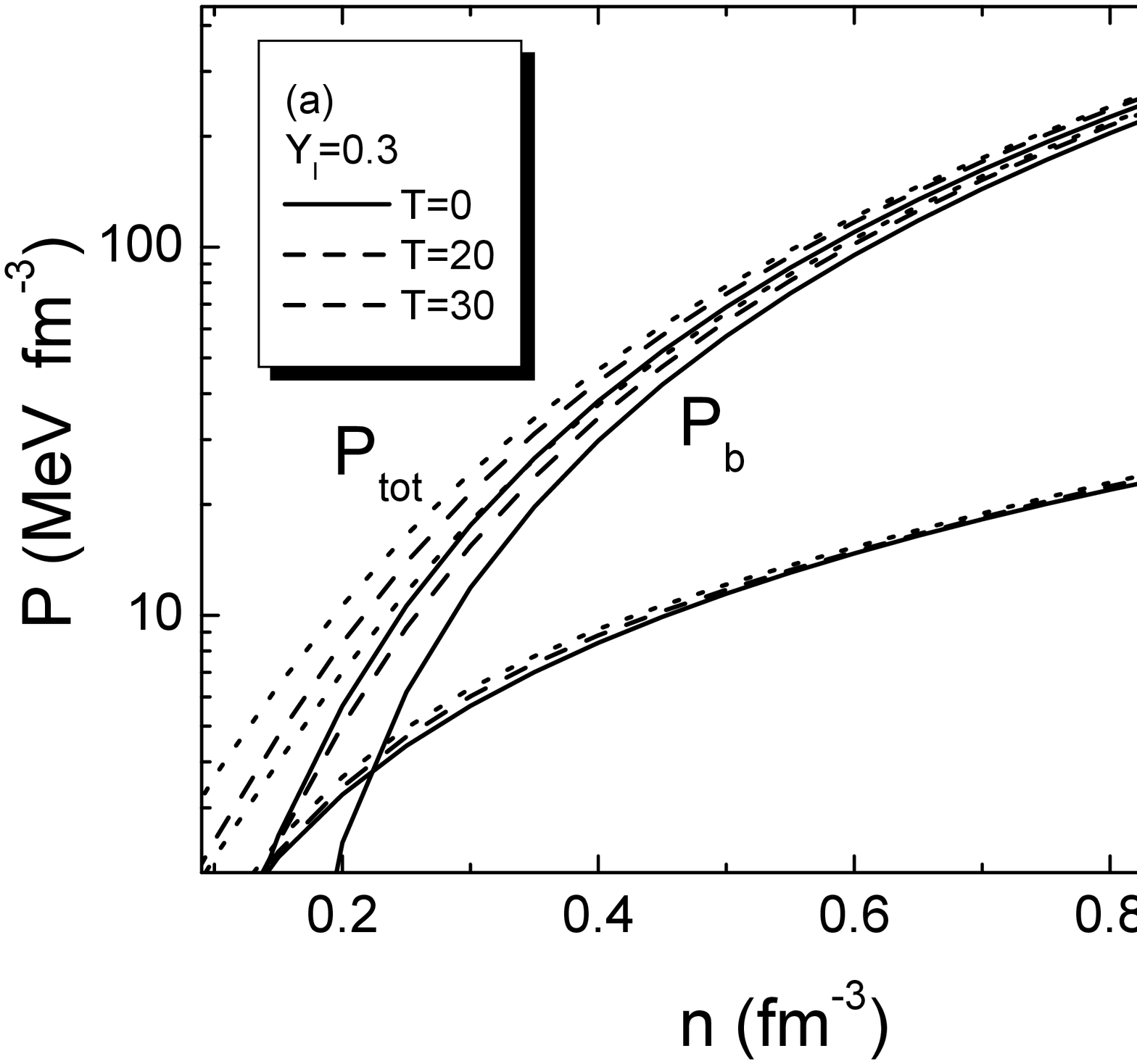}
\includegraphics[height=8.0cm,width=8.0cm]{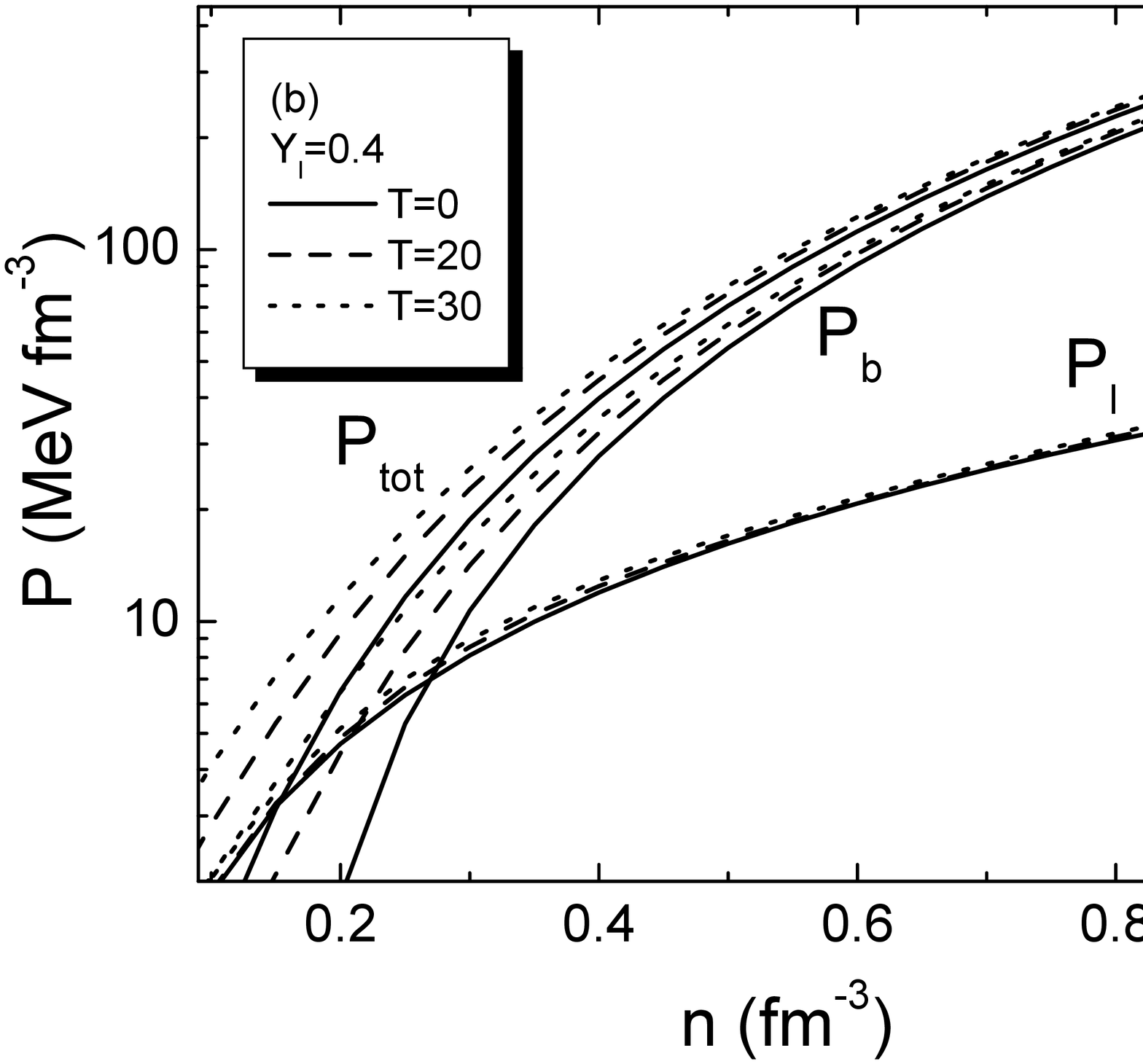}
\vspace{-3cm} \caption{The pressure of baryons ($P_b$), leptons
($P_l$) and the total pressure ($P_{tot}$) versus the baryon
density for various values of T for the cases a) $Y_l=0.3$ and b)
$Y_l=0.4$. } \label{}
\end{figure}
\begin{figure}
\centering
\includegraphics[height=8.0cm,width=8cm]{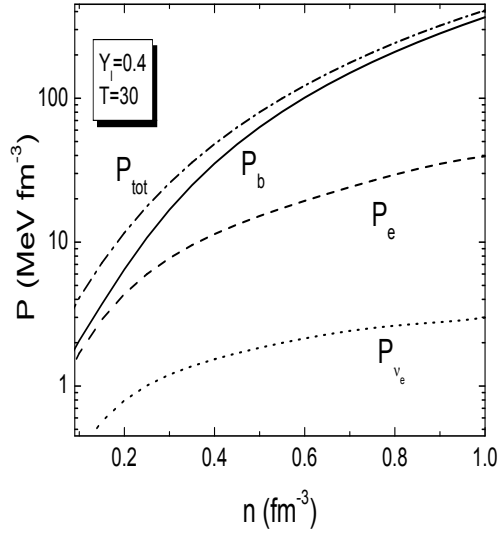}
\vspace{-3cm}
\caption{Pressures of respective components $P_i$ versus the
baryon density $n$ for the specific case with $T=30$ MeV and
$Y_l=0.4$. } \label{}
\end{figure}
\begin{figure}
\centering
\includegraphics[height=8.0cm,width=8cm]{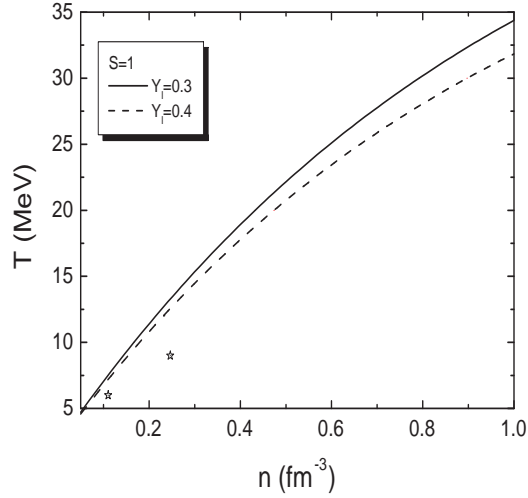}
\caption{Temperature $T$-density $n$ relation with $Y_l=0.3$
(solid line) and $Y_l=0.4$ (dashed line) for $S=1$. Stars denote
the results for the case with  $S=1$ and $Y_l=0.4$, extracted from
the results by Lattimer et al.~\cite{Lattimer-85}. } \label{}
\end{figure}
\begin{figure}
\centering
\includegraphics[height=10.0cm,width=7cm]{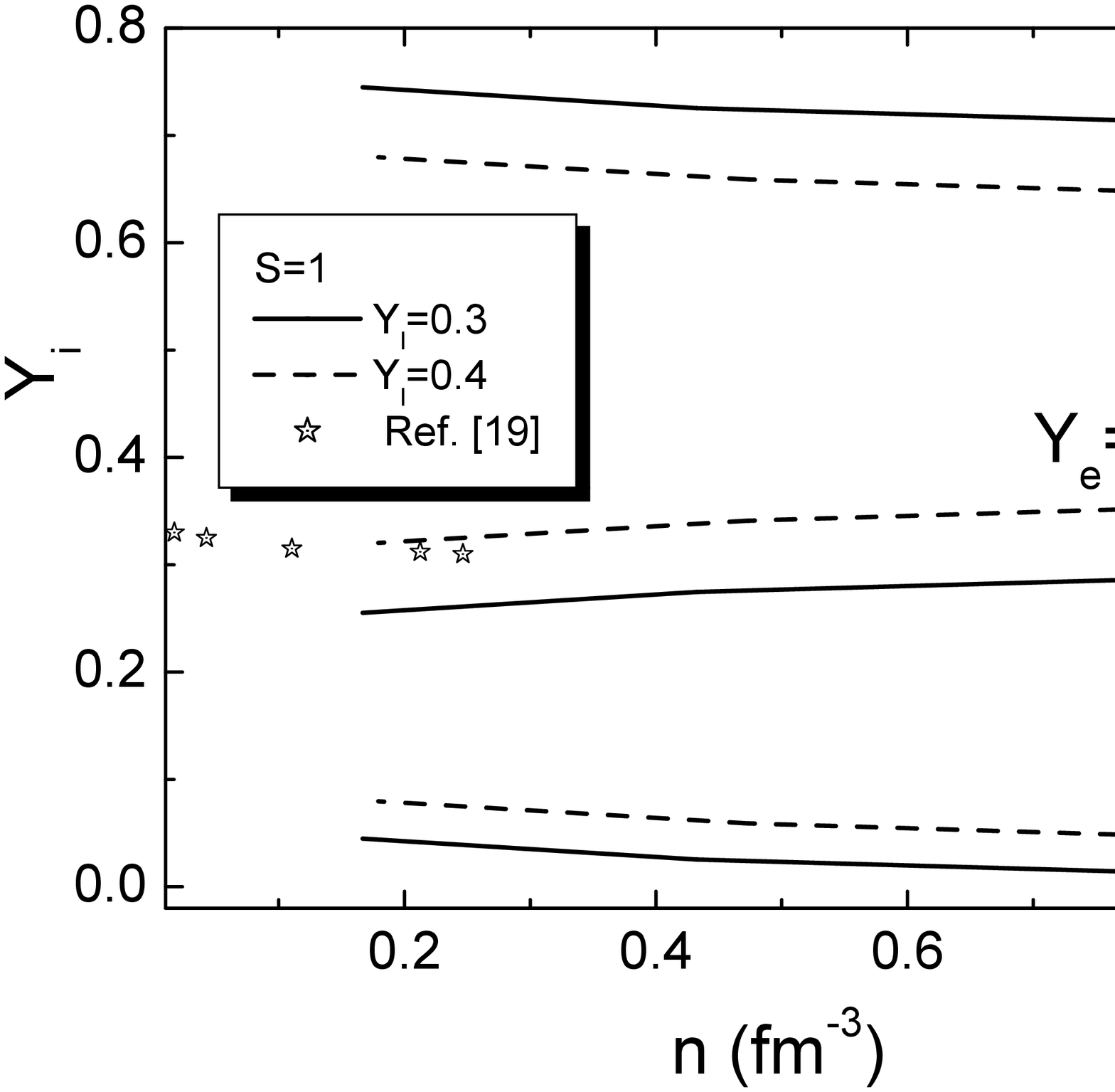}
\vspace{-4cm}
\caption{Fractions of respective components $Y_i$ as functions of
the density $n$ with $S=1$, for $Y_l=0.3$ (solid lines) and
$Y_l=0.4$ (dashed lines). Stars denote the results of $Y_p$ for
the case with  $S=1$ and $Y_l=0.4$, extracted from the results by
Lattimer et al.~\cite{Lattimer-85}. } \label{}
\end{figure}
\vspace{5cm}
\begin{figure}
\centering
\includegraphics[height=8.0cm,width=8cm]{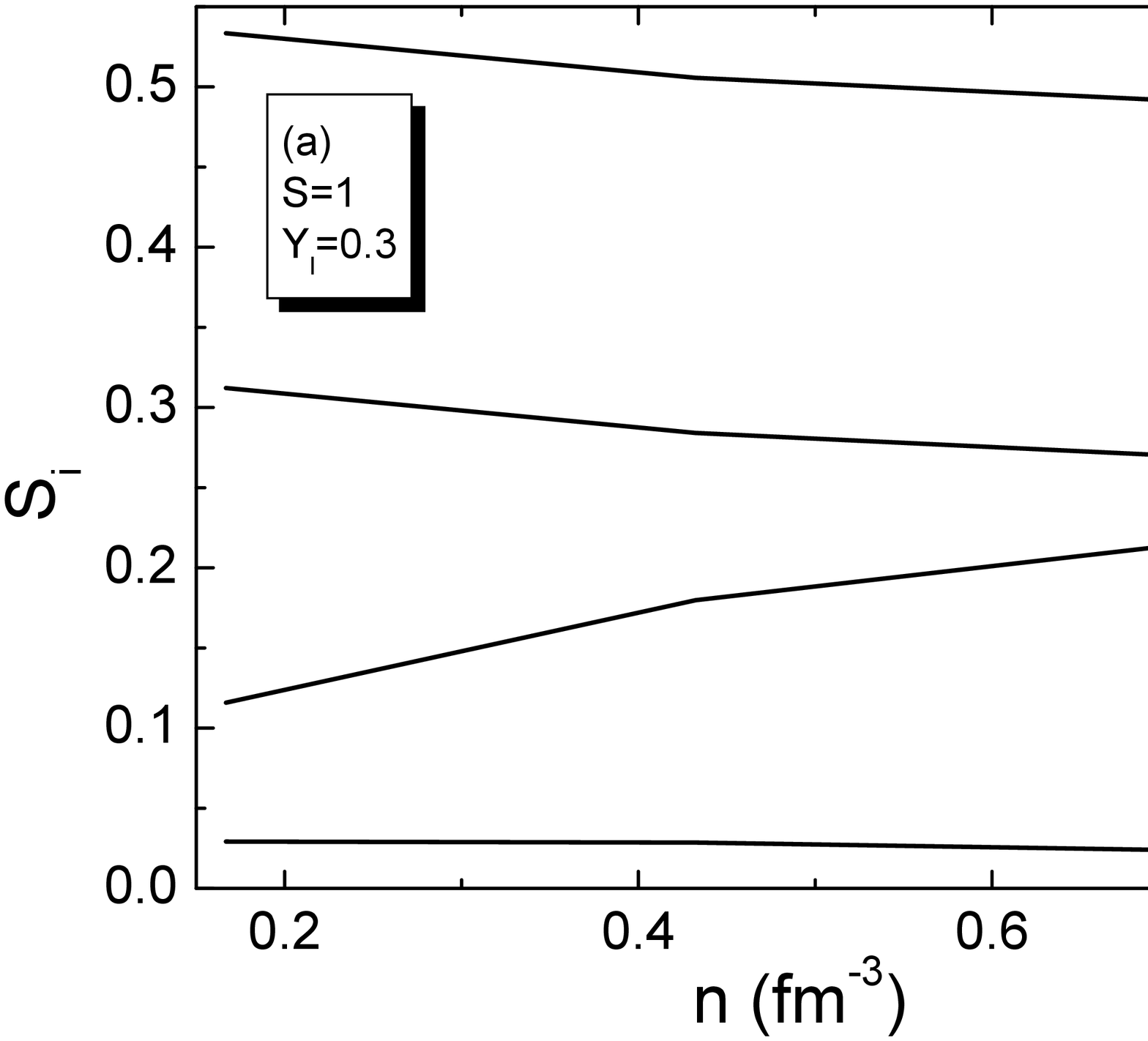}
\includegraphics[height=8.0cm,width=8.0cm]{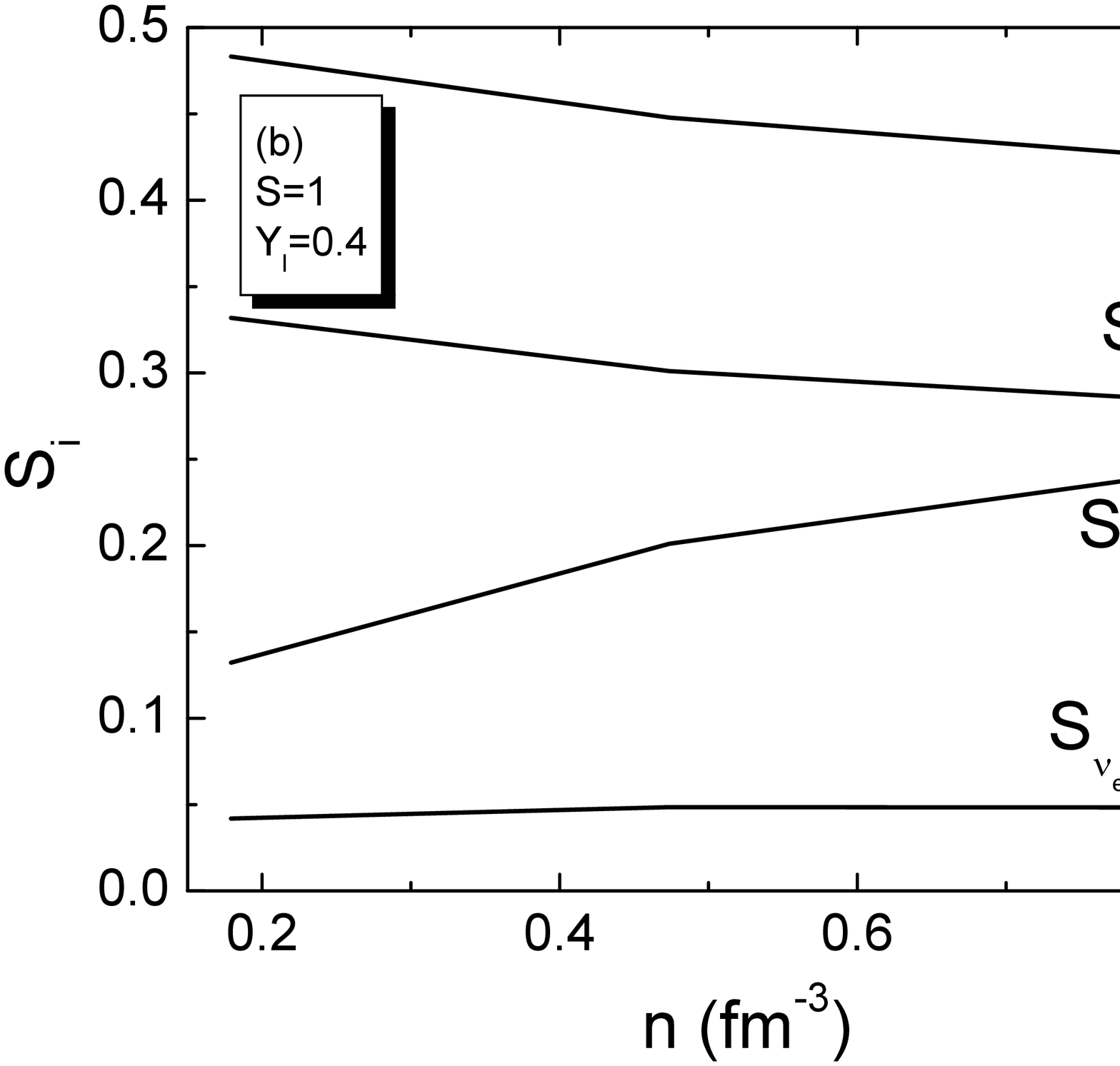}
\vspace{-3cm} \caption{Entropies per baryon of respective
components $S_i$ contributing to the total entropy $S=1$ of
supernova matter with total lepton fraction a) $Y_l=0.3$ and b)
$Y_l=0.4$, as functions of the density $n$. } \label{}
\end{figure}
\begin{figure}
\centering
\includegraphics[height=10.cm,width=7.cm]{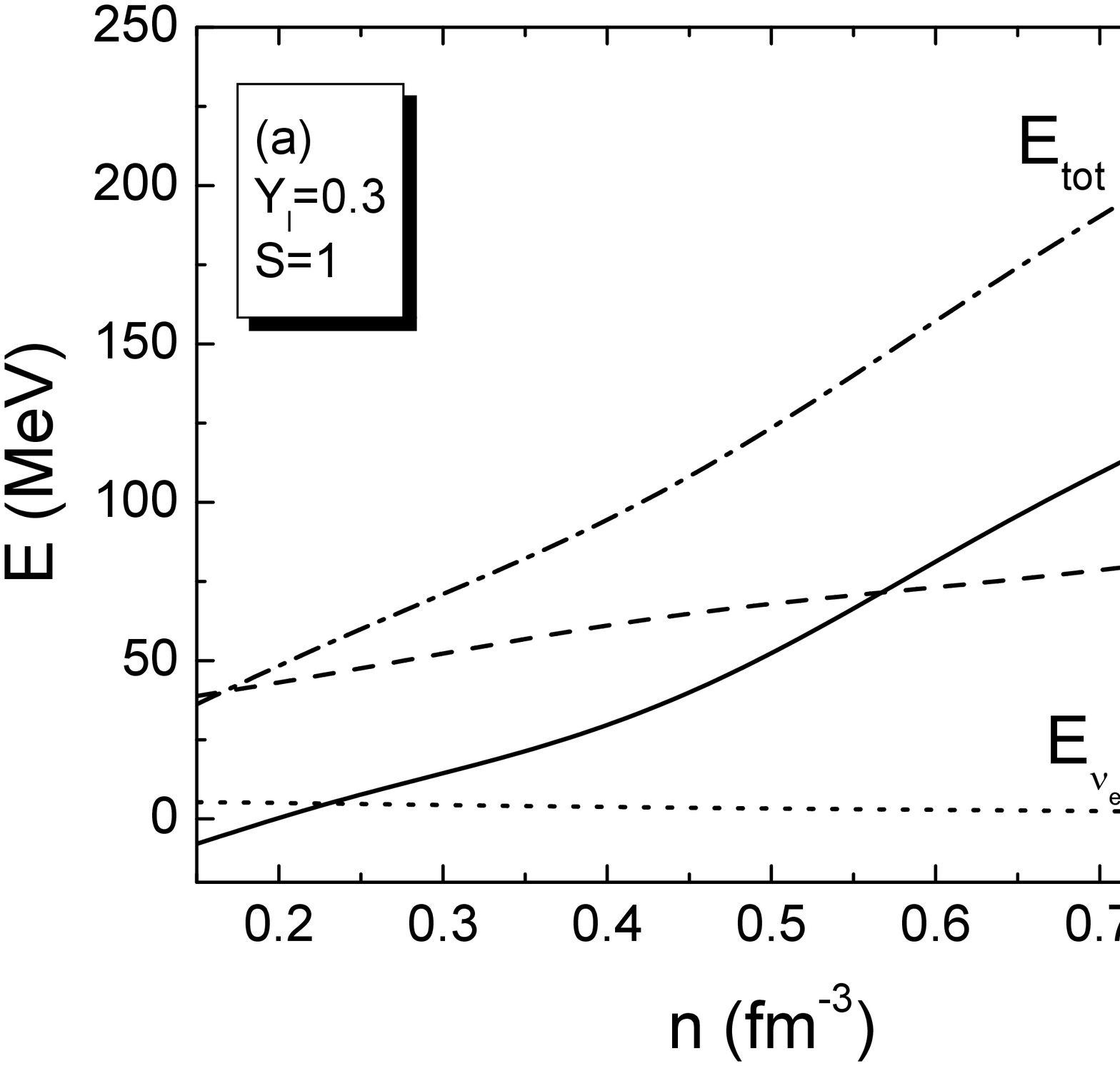}
\includegraphics[height=10cm,width=7cm]{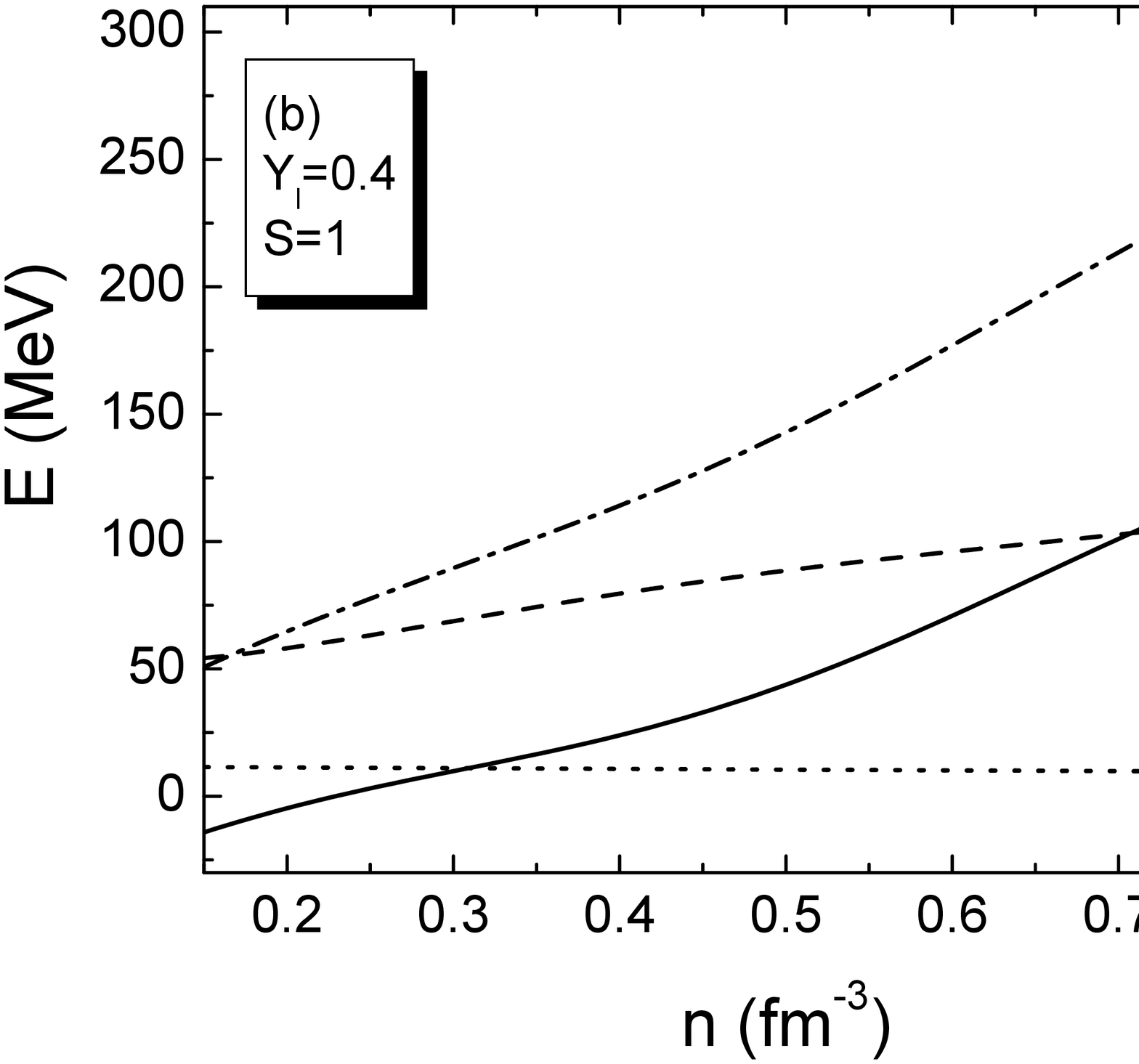}
\vspace{-1cm} \caption{Internal energies per baryon of respective
components $E_i$ versus $n$ with $S=1$ for the cases a) $Y_l=0.3$
and b) $Y_l=0.4$.} \label{}
\end{figure}
\
\begin{figure}
\vspace{4cm} \centering
\includegraphics[height=9.0cm,width=8cm]{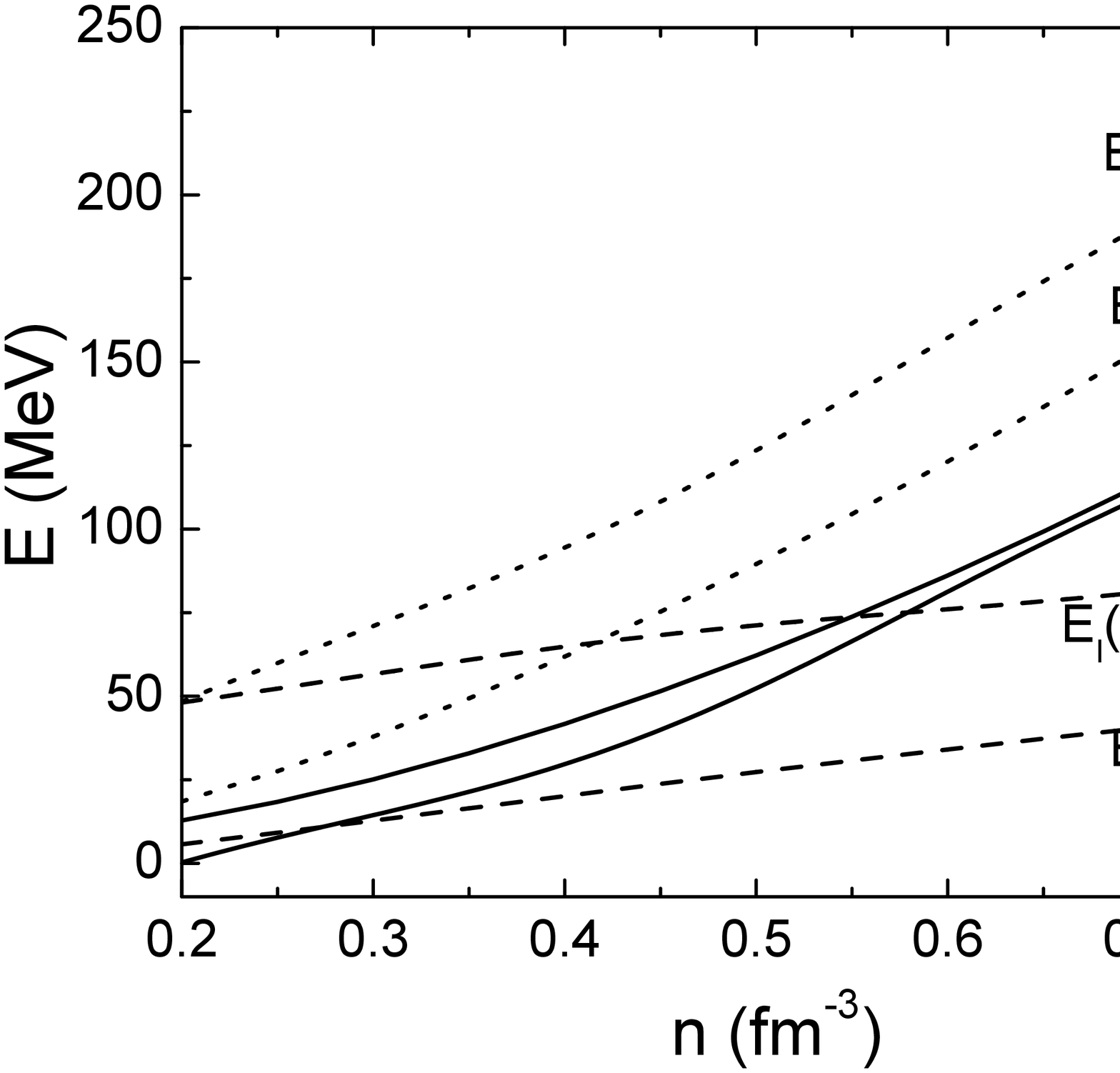}
\vspace{-4cm} \caption{Internal energy per baryon versus density
$n$ for dense supernova matter (SM) in comparison with that of
cold neutron star matter (NS) by applying the same model. The case
of supernova matter corresponds to $S=1$ and $Y_l=0.3$. The
contribution of each species is plotted separately.} \label{}
\end{figure}

\end{document}